\documentclass[trackchanges, twocolumn]{aastex701}

\usepackage{siunitx}
\usepackage[inkscapelatex=false]{svg}
\newcommand*{\arcminute}{\si{\arcminute}}
\newcommand*{\arcsecond}{\si{\arcsecond}}

\usepackage{gensymb}
\usepackage{amsmath}

\begin{document}

\title{Searching for Ultracool Dwarfs in Early LSST Data Products}

\author[orcid=0000-0003-1202-3683,sname=Honaker]{Easton J.\ Honaker}
\altaffiliation{Delaware Space Grant Graduate Fellow}
\affiliation{University of Delaware, Department of Physics \& Astronomy, Newark, DE 19716, USA}
\affiliation{University of Delaware, Data Science Institute, Newark, DE 19716, USA}
\email[show]{ehonaker@udel.edu}  

\author[orcid=0000-0002-8916-1972,sname=Gizis]{John E.\ Gizis} 
\affiliation{University of Delaware, Department of Physics \& Astronomy, Newark, DE 19716, USA}
\affiliation{University of Delaware, Data Science Institute, Newark, DE 19716, USA}
\email{gizis@udel.edu}

\author[orcid=0000-0003-2094-9128,sname=Aganze]{Christian Aganze} 
\affiliation{Kavli Institute for Particle Astrophysics \& Cosmology, Stanford University, Stanford, CA 94305, USA}
\affiliation{SLAC National Accelerator Laboratory, Menlo Park, CA 94025, USA}
\email{caganze@stanford.edu}

\author[orcid=0000-0001-5078-5457,sname=Chaini]{Siddharth Chaini}
\affiliation{University of Delaware, Department of Physics \& Astronomy, Newark, DE 19716, USA}
\affiliation{University of Delaware, Data Science Institute, Newark, DE 19716, USA}
\email{chaini@udel.edu}

\author[orcid=0000-0003-1953-8727,sname=Bianco]{Federica B.\ Bianco}
\affiliation{University of Delaware, Department of Physics \& Astronomy, Newark, DE 19716, USA}
\affiliation{University of Delaware, Data Science Institute, Newark, DE 19716, USA}
\affiliation{University of Delaware, Joseph R. Biden, Jr. School of Public Policy and Administration, Newark, DE 19716, USA}
\email{fbianco@udel.edu}

\author[orcid=0000-0001-6023-4974, sname={\v Z}erjal]{Maru{\v s}a {\v Z}erjal}
\affiliation{Instituto de Astrofísica de Canarias, E-38205 La Laguna, Tenerife, Spain}
\affiliation{Universidad de La Laguna, Dpto. Astrofísica, E-38206 La Laguna, Tenerife, Spain}
\email{maruska.zerjal@iac.es}

\author[orcid=0000-0002-1208-4833,sname=Mart{\' i}n]{Eduardo L. Mart{\' i}n}
\affiliation{Instituto de Astrofísica de Canarias, E-38205 La Laguna, Tenerife, Spain}
\affiliation{Universidad de La Laguna, Dpto. Astrofísica, E-38206 La Laguna, Tenerife, Spain}
\email{ege@iac.es}

\author[orcid=0000-0001-9273-5036,sname=Clarke]{Riley W.\ Clarke}
\affiliation{National University of Singapore, Department of Physics, 2 Science Drive 3, Singapore 117551}
\email{rclarke@nus.edu.sg}

\author[orcid=0009-0001-2333-7816,sname=Southwick]{Ashton Southwick}
\affiliation{University of Delaware, Department of Physics \& Astronomy, Newark, DE 19716, USA}
\email{ashtons@udel.edu}

\author[orcid=0009-0009-8636-7113,sname=Petrie]{Harrison Petrie}
\affiliation{University of Delaware, Department of Physics \& Astronomy, Newark, DE 19716, USA}
\email{hpetrie@udel.edu}

\author[orcid=0009-0003-6225-1964,sname=Blask]{Tyler Blask}
\affiliation{University of Delaware, Department of Physics \& Astronomy, Newark, DE 19716, USA}
\email{tvblask@udel.edu}

\correspondingauthor{Easton J.\ Honaker}
\begin{abstract}
The Vera C. Rubin Observatory’s Legacy Survey of Space and Time (LSST) promises to drastically accelerate the discovery of ultracool dwarfs (UCDs) over the course of its 10-year survey of the Southern Hemisphere. With the official start of LSST imminent, we showcase LSST's capabilities for discovering and characterizing UCDs using early commissioning data (Data Preview 1). The LSST photometric system at this stage remains poorly understood for faint UCDs. Thus, we begin by cross-matching Data Preview 1 against known UCD catalogs. We recover 1 known UCD from the Ultracool Sheet, 17 UCDs from the Dark Energy Survey, and 17 low mass stars from the Gaia Catalog of Nearby Stars. Using these known UCDs alongside recent spectroscopically-confirmed Euclid objects, we select 89 ultracool dwarf candidates in LSST fields, 17 of which are unique to this work. We present our candidates, a photometric temperature estimate, and discuss lessons learned from using early LSST data products. Finally, we turn to the future and predict potential UCD counts in upcoming LSST commissioning data (Data Preview 2), which is expected to be available to the Rubin community in 2026. Using synthetic populations of brown dwarfs, we forecast over 17,000 objects may be discovered and characterized in Data Preview 2. We predict that several hundred known objects and thousands of as-of-yet undiscovered UCDs may be detected in Data Preview 2 fields. 
\end{abstract}


\section{Introduction}

The long wait is over: after decades of preparation and anticipation, the Vera C. Rubin Observatory's Legacy Survey of Space and Time (LSST) has arrived. The survey will scan $\geq18,000$ square degrees of the Southern sky over its 10 year timeline, revealing the ``greatest cosmic movie'' ever made of over 17 billion stars and 20 billion galaxies \citep{lsst2019}. LSST will exceed previous optical survey depths by several magnitudes, capture millions of temporal variations, and reach millimag and milliarcsecond photometric and astrometric precisions, respectively. Science cases from all corners of the astronomical community will benefit from the images and catalogs Rubin Observatory will deliver. 

The proper LSST survey is scheduled to begin mid 2026, with the transient ``Alert stream'' \citep{RTN-011} already started on February 24, 2026. The survey design and observation scheduling was created to enable as many scientific use cases as possible while also maximizing synergy with other contemporary observatories \citep{2022ApJS..258....1B}.
Rubin Observatory released its commissioning data as Data Preview 1 (DP1), in July 2025 \citep{dp1full}. This commissioning data provided the first opportunity for users to understand the LSST data processing pipelines and products. Already, recent studies showcase LSST's potential across many fields of astronomy (e.g. \citealt{carlin25_dp1, Chandler25_dp1,choi25_dp1,Cordoni25_dp1,Romanowsky25_dp1,Wang25_dp1, greenstreet26_dp1}). In this work, we focus on our coolest neighbors and how LSST can contribute.

Ultracool dwarfs (UCDs, \citealt{Kirkpatrick05}) are objects with effective temperatures $T_{\text{eff}}\leq 3,000$ K; this category contains both very-low-mass stars (mid- to late-M dwarfs) as well as brown dwarfs (L, T, and Y dwarfs).\footnote{While UCD is a larger umbrella term and refers to more than just brown dwarfs, for the purposes of this paper the terms are used interchangeably except for cases specifically referring to the exclusion or inclusion of M dwarfs.} UCDs are intrinsically faint objects ($<10^{-4} L_\odot$, \citealt{Burrows01}) that emit the majority of their flux in the infrared. As such, the majority of known UCDs, specifically brown dwarfs, are found in the closest portion of the Milky Way, our Solar Neighborhood ($d\lesssim200$ pc). 

Brown dwarfs are substellar objects below the hydrogen fusing mass limit that perpetually cool over their lifetime \citep{Kumar}. The transitions from warmer M spectral types ($T_{\text{eff}}>2,400$ K) to L types ($T_{\text{eff}}\approx 2,400$ K to $1,400$ K) is marked by the weakening of optical TiO bands from dust condensation, significant H$_2$O absorption features shaping the objects' spectra, and photometric reddening \citep{Jones97, kirkpatrick99, martin99}.  Once objects cool to $\sim1,200 - 1,400$ K, they reach the L/T transition, where the dusty clouds sink deeper into the atmosphere, resulting in a clearing of optical depth and bluer photometric colors \citep{sm08}. The exact mechanisms and timescales associated with this transition from cloudy to cloudless atmospheres is an open research question with ongoing modeling work. The impact of a large sample of objects at the L/T transition from LSST on this research cannot be overstated. LSST will constrain the brown dwarf luminosity function at the L/T transition and extend recent work on the local mass and luminosity functions \citep{Kirkpatrick2024}. Lastly, as objects cool further to the T/Y transition ($T_{\text{eff}}\lesssim 600$ K), icy water clouds dominate their spectra and disequilibrium chemistry becomes critical \citep{morley14}. Since LSST is an optical instrument, it will likely be unable to detect objects at the T/Y transition, unless they are exceptionally nearby. Nevertheless, LSST's photometric capabilities promise to reveal a myriad of M, L, and early T dwarfs. As LSST progresses, one will be able to assemble an unprecedentedly large sample of brown dwarfs to answer some of the field's biggest questions with statistical confidence.

The appeal of studying UCDs with LSST extends beyond just the brown dwarf community. Since brown dwarfs cannot support hydrogen fusion, they retain the metallicity information from their formation site, making them excellent probes of galactic archaeology and evolution \citep{Barbara12, Ryan17, martin18_Li, Aganze22}. Similarly, these objects have high proper motions and their kinematics can be used to study Galactic structure and dynamical heating \citep{Faherty_BDKP1,dupuy17,best24}. Many brown dwarfs share similar temperatures and atmospheric chemistry with exoplanets, allowing for the study of exoplanet-analogous atmospheres without contamination from a host star \citep{Faherty16_analogs}. For the extragalactic community, a better understanding of UCD selection techniques can enhance their sample purity since ultracool dwarfs' red photometric colors are common contaminants in high redshift quasar samples and easily confused for little red dot galaxies \citep{ceers, cosmos25, Hainline25}. LSST's multi-band temporal coverage and alert stream can also aid variability and flare studies and trigger follow-up observations \citep{Gizis13, Schmidt16}. The expected astrometric performance of LSST, with a sufficient baseline, can act as an extension of the Gaia mission for faint objects, enabling proper motion and astrometric binary studies \citep{lsstsciencebook}.

All of these science cases begin with the detection, identification, and characterization of UCDs. Historically, the discovery of UCDs has been driven by wide-area photometric sky surveys like the Two Micron All Sky Survey (2MASS,  \citealt{2mass,kirkpatrick99, gizis20002mass}), Pan-STARRS1 (PS1, \citealt{ps1,Deacon14,best18ps1}), the Wide-field Infrared Survey Explorer (WISE, \citealt{wise,Cushing11Y,Kirkpatrick2011Wise}), and the Sloan Digital Sky Survey (SDSS, \citealt{sdss,west04}). The next generation of instruments and surveys, LSST, Euclid \citep{euclidmission}, JWST \citep{JWSTmission}, SphereX \citep{spherexmission}, and Roman \citep{romanmission}  have already (or soon will) begin to scan the skies.

Many of the science cases previously mentioned strongly benefit from spectroscopic follow-up. JWST is exceptionally good at spectral characterization of UCDs and has revolutionized our understanding of the coldest T and Y dwarf atmospheres (e.g. \citealt{beiler23JWSTy,Luhman24_coldbd,burgasser24,bardalez25_TYbin}) as well as the atmospheric structures in warmer L dwarfs (e.g. \citealt{vos23,manjavacas24, phillips24, patapis25}). JWST is the premier follow-up tool but is best suited to pencil-beam type surveys, not the wide field surveys that discover the majority of UCDs. Follow-up targets are required to have precise parallaxes and proper motions; cases with substandard astrometry can lead to missed targets and failed observations, making accurate measurements from larger photometric surveys is imperative. The actual discovery and identification of objects that are well-suited for follow-up must come from LSST, Euclid, and Roman. LSST uniquely offers optical wavelength and temporal coverage of the Southern sky with hundreds of observations of every patch of sky over the course of the 10-year survey.

This paper presents a showcase of LSST's early data products and their capabilities to enable the discovery of UCDs. We divide our exploration into sections aligned with the past, present, and future. First, we describe Data Preview 1 in Section \ref{sec:dp1}. In Section \ref{knownucd}, we explore previously known UCDs and their properties within LSST DP1. Next, we consider the present state of UCD searches in Section \ref{sec:candidateselection} by selecting UCD candidates using traditional color cut methods. In Section \ref{future} we look to future LSST data products and predict UCD counts and highlight key considerations to fully utilize LSST's capabilities. Lastly, we summarize our conclusions in Section \ref{conclusion}.

\section{LSST Data Preview 1}\label{sec:dp1}

In July 2025, the Rubin observatory released its first set of commissioning data as Data Preview 1 (DP1, \citealt{dp1full}), giving the Rubin community its first look at on-sky LSST data products. DP1 included processed images and catalogs of seven $\sim 1$ square degree fields with varying filter combinations and depths observed using the LSST commissioning camera, LSSTComCam. In DP1, outputs from source extraction run on individual exposures, known as visits, are collected in the Source tables \citep{dp1Source}, which are band-specific and ideal for transient objects. For static science cases, detected objects from deep, coadded images are reported in the Object table \citep{dp1Object}. Entries in the Object table are detections above a signal-to-noise ratio of 5 in at least one coadded band and are a collection of astrometric and photometric measurements over many visits. Unique to the Object table, some measurements, such as the object position and extendedness, are available for each band individually but also for the ``reference band'' used by the coaddition pipelines. The reference band is determined by detection significance for a source and preferentially adopted in $i,r,z,y,g,u$ order. Measurements on other bands, such as \texttt{*\_psfMag} or morphology calculations, are therefore generally forced measurements at positions determined by the reference band. A more in depth discussion of the Object catalog and the pipelines that produce it can be found in \citet{PSTN-019}. Since DP1 includes both Object and Source tables based on deep coadds and individual images, respectively, we will mirror this nomenclature by referring to entries in the Object table as objects and entries in the Sources table as sources. In this work, we exclusively use the Object table and access all DP1 data products through the Rubin Science Platform (RSP).\footnote{\url{https://data.lsst.cloud/}}

One of the first challenges with new observatories is characterizing the photometric system and data products. We begin by searching for known UCDs within DP1. We focus our efforts on four of the seven fields: Extended Chandra Deep Field South (ECDFS, $\alpha,\delta$: $53.13, -28.10$), Euclid Deep Field South (EDFS, $\alpha,\delta$: $59.10, -48.73$), Low Ecliptic Latitude field (LEL, $\alpha,\delta$: $37.86, 6.98$), and Low Galactic Latitude field (LGL, $\alpha,\delta$: $95.00, -25.00$).  We do not prioritize the remaining three fields because the 47 Tuc Globular Cluster field lacks $z$ band photometry and significant crowding limits the LSST analysis pipelines, the Fornax Dwarf Spheroidal Galaxy observations contains no $z$ or $y$ band photometry, and, lastly, the Seagull Nebula field observations are significantly shallower than all other fields and include no $i$ or $y$ band photometry. 

\section{Known UCDs in DP1}\label{knownucd}

We cross-match DP1 with other surveys' counterparts using the Object table \citep{dp1Object}. We use photometry in the $r$, $i$, $z$, and $y$ band (as available, LEL lacks $y$ band photometry), avoiding $u$ and $g$ bands. We primarily avoid $u$ band since UCDs emit very limited flux at such short wavelengths. The LSSTComCam CCDs (used to generate DP1) have a known red leak in the $g$ band filter range, leading to unphysically bright $g$ magnitudes for cooler objects.\footnote{For a detailed discussion on the $g$ band red leak, see SITCOMTN-152 \citep{SITCOMTN-152}.} As such, we also avoid using the $g$ band in DP1. The red leak is unique to LSSTComCam and is not expected to occur with the actual survey camera, LSSTCam, in future data releases.

\subsection{The Ultracool Sheet}\label{sec:UltracoolSheetmatches}

We first search for known UCDs from the Ultracool Sheet \citep{ucs} in all six DP1 fields. Only one object, ULAS J023144.49+063602.2 (hereafter ULAS J0231+0636, phototype T2.8, \citealt{Cardoso15}), falls within any of the DP1 regions. We identify it as LSST-DP1-O-648364758310848043 in the LEL field, $4.2\arcsecond$ away from its J2000 position in the Ultracool sheet. ULAS J0231+0636 was discovered using methane imaging by \citet{Cardoso15} and was previously included in UKIDSS-DR8 \citep{Lawrence12} and later in CatWISE2020 \citep{Maroc21}. With $Y_{\text{MKO}} = 20.07\pm0.18$ \citep{Lawrence12} and $J_{\text{MKO}} = 18.98\pm0.11$ \citep{Cardoso15}, ULAS J0231+0636 falls well below Gaia's sensitivity limit; additionally, ULAS J0231+0636 is not within the Dark Energy Survey's footprint. This detection by LSST is, to the best of our knowledge, the first optical observation of ULAS J0231+0636.

\subsection{Dark Energy Survey}\label{sec:DESmatches}
\begin{figure*}
    \centering
    \includegraphics[width=\textwidth]{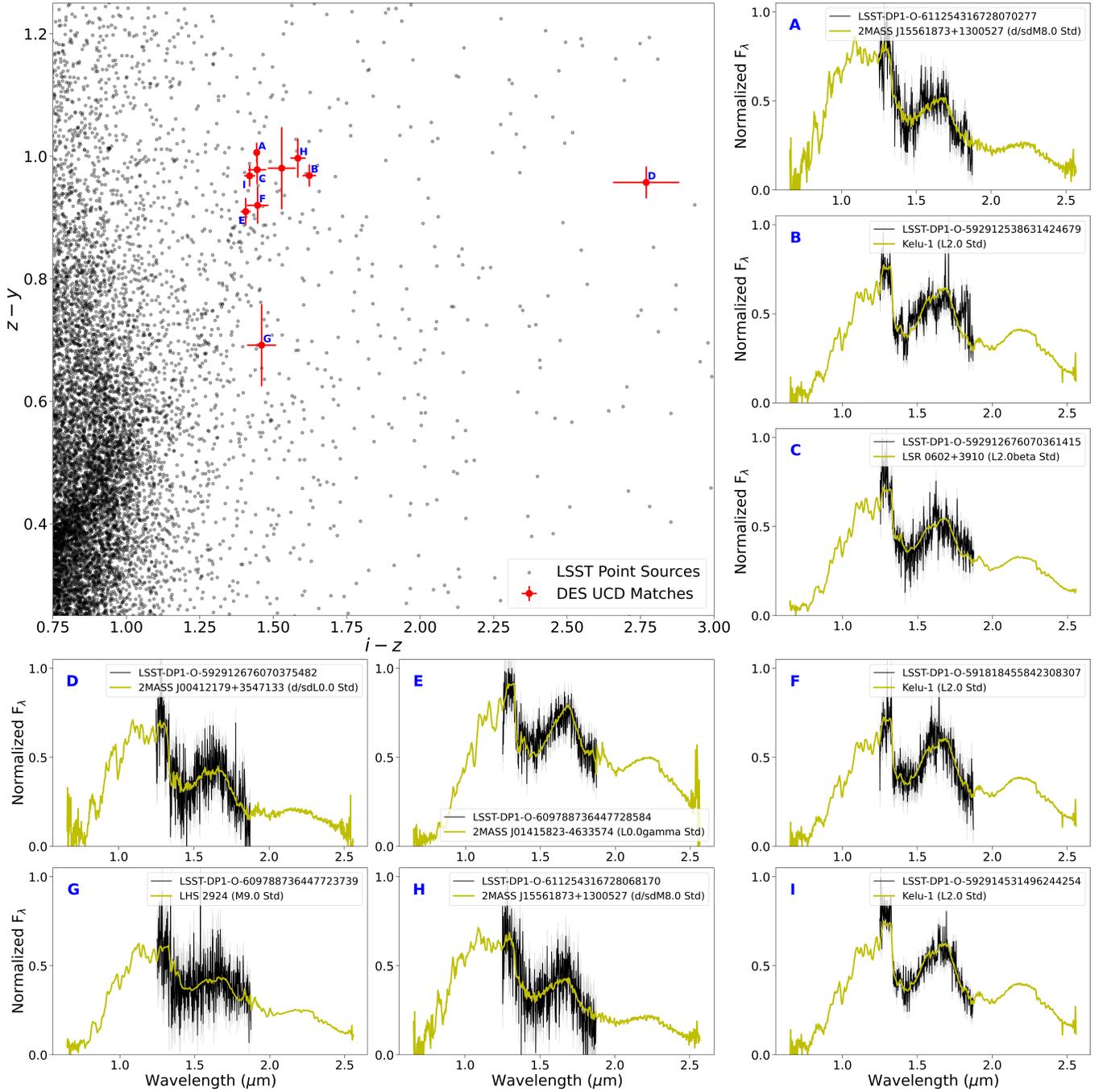}
    \caption{The DES candidates with LSST counterparts are shown in red against LSST point sources in the upper left panel on a LSST color-color plot. For the nine cross-matched counterparts with full $i,z,y$ coverage and Euclid spectra with median $\textrm{SNR}>3$, we display each spectrum in black with the best-fitting spectral standard in yellow. LSST-matched candidates and their spectra are paired using labels A-I.} 
    \label{fig:DESmatches}
\end{figure*}

In the Dark Energy Survey's Data Release 2 (DES DR2), \citet{dalponte23} identified 19,583 UCD candidates across the 5,000 deg$^2$ footprint using phototype and template matching techniques. The DES footprint overlaps three of the $\sim$1 square degree LSST DP1 fields: ECDFS, EDFS, and Fornax. We search for counterparts to their candidates in LSST using a $5\arcsecond$ search radius and recovered 17 of the DES candidates (7 in ECDFS, 5 in EDFS, and 5 in Fornax). For each unique DES candidate, we visually inspected cross-matched LSST objects to determine the single best LSST counterpart. In all cases, the best counterpart was also the closest cross-match; the mean separation between the DES candidate positions and their LSST counterparts is $0.175\arcsecond\pm 0.150\arcsecond$. The majority of offsets are smaller than the instruments' respective pixel scales (LSSTComCam: $0.2$\arcsec/px, DECam: $0.2637$\arcsec/px). We list the 17 pairings' DES coadd identifiers, LSST DP1 Object table identifiers, and their positional offsets in Table \ref{tab:DESmatches}.  Of the 12 objects that are in the ECDFS and EDFS fields, we show 10 of their positions in a color-color diagram in Figure \ref{fig:DESmatches}. The remaining two are only detected in one of the three $i,z,y$ bands and we cannot construct the relevant colors. Objects in the Fornax field lack $z,y$ photometry and in most cases are too faint to create reliable colors using $u,g,r$. Thus, they are not included in Figure \ref{fig:DESmatches}. For the 10 objects shown, the mean differences in their $i,z,y$ band magnitudes (LSST - DES) and associated standard deviation are $\Delta i = 0.475\pm0.410$ mag, $\Delta z = 0.498\pm0.042$ mag, and $\Delta y = 0.074\pm0.076$ mag.\footnote{Both DES and LSST use AB magnitudes. The respective $i,z,y$ bandpasses are similar but not identical in transmission.} The $i$ band differences are skewed by the red outlier LSST-O-DP1-592912676070375482 (also further discussed in Section \ref{sec:candidates}). Removing this object reduces the magnitude offset in the $i$ band to $\Delta i=0.386\pm0.252$ mag. In Section \ref{sec:candidates}, we select UCD candidates in the ECDFS and EDFS fields. All 10 DES candidates in the ECDFS and EDFS fields with multi-band detections make our final candidate selections.

For the 10 DES objects in the ECDFS and EDFS fields with $i,z,y$-based colors, we can validate the phototypes from \citet{dalponte23} using Euclid spectra. All 10 objects are phototyped as either L0 or L1 dwarfs. We retrieve all 10 objects' spectra by submitting TAP queries through the IRSA Euclid archives. For each retrieved spectrum, we discard flux bins where mask values are odd or values $\geq64$, retaining only Euclid-recommended points and  then computed the median signal-to-noise ratio (SNR) from $1.3$ $\mu$m to $1.8$ $\mu$m. Nine of the ten retrieved spectra have SNR $>3$; we remove the one spectrum below our threshold (LSST-DP1-O-611255691117613253) from our analysis. We spectral type each of the remaining nine spectra using the SPLAT software package \citep{splat} and the infrared standards within via $\chi^2$ minimization. For each object, we compare first to standard dwarf templates and then to  youth and subdwarf templates to find the lowest $\chi^2$ result before visually inspecting all final results. The resulting best fits are shown in the side panels of Figure \ref{fig:DESmatches} for each object. 

We now briefly discuss the resulting spectral types for the 9 DES-LSST counterparts. Four of the nine objects are best fit by standard dwarf templates, three by lower metallicity dwarf/subdwarf templates, and the remaining two by low-gravity templates. In all cases, resulting spectral types range from M8 to L2. We note that the Euclid spectra only extend from $1.2-1.8$ $\mu$m, which contains the entire $H$ band region but only part of the $J$ band and none of the $Y$ or $K$ band sections of an object's spectral energy distribution (SED). For late-M and early-L dwarfs, the $0.9 -1.4$ $\mu$m region is critical to spectral typing \citep{Kirkpatrick10}. While we can broadly determine the best spectral type from the available Euclid spectra, more complete coverage is necessary to confirm these objects' nature. For example, while a triangular $H$ band shape is a diagnostic for young, low-gravity objects, a more complete spectrum that allows for the calculation of gravity-sensitive indices from \citet{Allersgrav} is necessary to confirm LSST-DP1-O-609788736447728584 and LSST-DP1-O-592912676070361415 as young, low-gravity objects. We note that LSST-DP1-O-591818455842308307 (Panel F), is best fit by the L2 standard Kelu-1, but has additional flux from $1.55-1.65$ $\mu$m. We attempted to fit this object using low-gravity, subdwarf, and binary templates but no single object or combined binary template offers substantial improvement in fitting statistics or visually fits better than the L2 standard. As such, we note this object may benefit from additional follow-up observations to understand the flux excess.

\begin{deluxetable}{c c c}
    \tablecolumns{3}
    \label{tab:DESmatches}
    \tablecaption{DES and LSST UCD counterparts}
    \tablehead{\colhead{DES Coadd Id} & \colhead{LSST objectId} & \colhead{Offset}  \\ \colhead{}  & \colhead{LSST-DP1-O-} & \colhead{(arcsec)} }
    \startdata
        1328918880 & 605447177346486023 & 0.112 \\
        1328887428 & 605447864541253066 & 0.040 \\
        1328887563 & 605447795821799979 & 0.210 \\
        1329176716 & 604070726227491372 & 0.364 \\
        1329107252 & 604071344702753879 & 0.002 \\
        \hline
        1389041770 & 611256653190268598 & 0.053 \\
        1394561870 & 611255691117613253 & 0.191 \\
        1394602941 & 611254316728070277 & 0.076 \\
        1395307524 & 611254316728068170 & 0.185 \\
        1395334859 & 609788736447723739 & 0.067 \\
        1395327167 & 609788736447728584 & 0.144 \\
        1399577467 & 611253492094352857 & 0.144 \\
        \hline
        1425205963 & 592912676070375482 & 0.151 \\
        1425208507 & 592912676070361415 & 0.376 \\
        1431403950 & 591818455842308307 & 0.162 \\
        1431388259 & 592912538631424679 & 0.626 \\
        1432755994 & 592914531496244254 & 0.076 \\
    \enddata
    \tablecomments{Listed cross-matches are ordered by LSST RA. The first 5 entries are in the DP1 Fornax field, the following 7 are in ECDFS, and the final 5 are in EDFS.}
\end{deluxetable}

\subsection{Gaia Catalog of Nearby Stars}\label{sec:GCNSmatches}
In addition to UCDs from the Ultracool Sheet and the DES DR2 candidate lists, we search for known UCDs from the Gaia DR2 catalog of UCDs \citep{gaiadr2ucdsI}. However, none of the 321 Gaia DR2 UCDs overlapped the four DP1 regions of interest (ECDFS, EDFS, LEL, LGL). We extended our search to include known low mass objects from the Gaia EDR3 Catalog of Nearby Stars (GCNS, \citealt{GaiaGCNS}). The GCNS is comprised of vetted objects within 100 pc and includes main sequence stars as well as a few UCDs. However, UCDs are generally too faint to be detected with Gaia making the GCNS incomplete for these spectral types. 

From the GCNS, we propagated positions from the epoch 2016.0 coordinates in Gaia EDR3 to an average visit epoch of 2025.0 for DP1 using all available Gaia proper motion, parallax, and radial velocity information. Using TOPCAT, we cross-matched the propagated positions of all GCNS objects within 1.25$\degree$ of the DP1 field centers with the LSST Object catalog. Since the Object catalog is an averaged position over many images in multiple bands per object, we cross-matched using a $3\arcsecond$ search radius to account for astrometric precision and coaddition averaging differences. We also cross-matched using a $5\arcsecond$ radius but found little to no improvement over the $3\arcsecond$ radius.

In total, we found potential cross-matches for 40 of the GCNS objects: 19 GCNS objects in LEL, 9 in ECDFS, 6 in LGL, and 6 in EDFS. Using the GCNS objects' absolute Gaia $G$ magnitude, $BP-RP$ color, and Gaia reported temperature (as available), we approximated broad spectral types for the 40 objects. In the sample, there are 1 F-type star, 1 G-type star, 7 K-type stars, 28 M dwarfs, and 3 white dwarfs. Based on absolute $G$ and $RP$ relations from \citet{gaiaucdsII}, all M dwarfs are warmer than M8. 

These 40 GCNS objects matched to 134 LSST objects; visual inspection revealed that 18 GCNS objects had reliable 1:1 matches, but the remaining 22 GCNS objects were each matched to three or more LSST objects. All objects with clean LSST matches were M dwarfs or white dwarfs while nearly all objects with more than 3 potential LSST counterparts were K or earlier spectral types. Further reviewing the matches reveals that for nearby, early-type objects, namely G, K, and early M dwarfs, the objects are saturated in LSST.\footnote{From the LSST Science book \citep{lsstsciencebook}, a 15 second exposure with $0.7\arcsecond$ seeing will saturate at  $u, g, r, i, z, y$ = $14.7,\space 15.7,\space 15.8,\space 15.8,\space 15.3,$ and $13.9$ mag, respectively.}  In the current LSST processing pipelines, nearby saturated stars cover a large enough angular size to be detected as an extended object, deblended into multiple peaks, and ultimately cataloged as several ``child'' objects. These reported ``children'' from the deblending processing appear throughout the wings and center of saturated objects but cannot provide a single definitive position or magnitude measurement for the original, saturated source. In fact, these saturation artifacts are color outliers that can easily be confused with extreme-color objects and contaminate color-color selection of UCDs. In DP1, there is no current flag system in place to warn that such deblending artifacts are near a known bright Gaia source. Similarly, there currently is no mechanism to perform astrometric or photometric measurements on the original saturated sources within the LSST pipelines. It may be possible to reconstruct the saturated object’s LSST position using its deblended children, but that is beyond the scope of this work. We conclude that studies interested in low mass stars should focus on objects beyond the GCNS 100 pc limit to avoid this issue. 

\begin{figure}[ht!]
    \centering
    \includegraphics[width=0.5\textwidth]{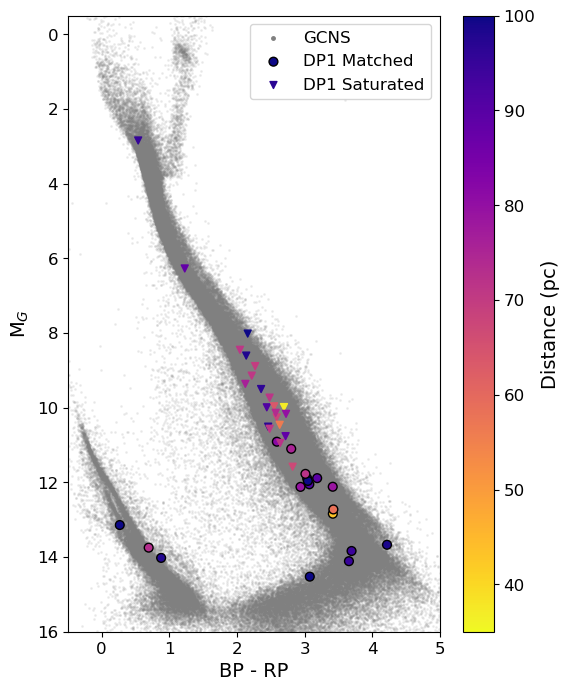}
    \caption{Cross-matched GCNS and LSST DP1 objects are shown on top of the entire GCNS (gray dots) on a Gaia $M_G$ vs $BP-RP$ color-magnitude diagram. GCNS objects with LSST cross-matches are color-coded by distance. Triangles denote LSST saturated objects whereas circles represent resolved LSST counterparts. Nearly all low mass main sequence objects with $M_G < 12.0$ within 100 pc become saturated in LSST.}
    \label{fig:saturatedcmd}
\end{figure}

For the 18 GCNS objects that did not saturate in the LSST exposures, we tabulate their LSST astrometry and photometry in Table \ref{tab:GCNSLSST}. The remaining saturated objects are also listed in Table \ref{tab:GCNSLSST}, but lack LSST-specific information. Positions in Table \ref{tab:GCNSLSST} are from LSST for non-saturated matches and from Gaia EDR3 for saturated sources. In Figure \ref{fig:saturatedcmd} we show the positions of all 40 objects on a Gaia color-magnitude diagram with points color-coded by distance. Most GCNS objects lie $60 - 100$ pc away while three objects are between $35 - 60$ pc. Saturation is a result of both distance and spectral type. Objects with $M_G \geq 12.0$ are generally unsaturated within 100 pc whereas all objects with $M_G < 12.0$ are saturated in DP1.  To further investigate saturation limits, we performed synthetic photometry on BT-SETTL models \citep{btsettl} to estimate how far an object must be to avoid saturating in LSST. We use 6,000 K, 5,000 K, and 4,000 K models to model F, G, and K stars, respectively. Assuming the objects first saturate the shallowest band, the $y$ band, we compute a limiting distance by translating LSSTComCam $g$, $r$, and $i$ band synthetic photometry to Gaia $G$, $BP$, and $RP$ using DP1 photometric conversions \citep{RTN-099}. Then, we assume an absolute Gaia $G$ magnitude of $M_G =4.0,\space 6.0,\space 7.5$ for the F, G, and K star models, respectively. We select these values as broad approximations for a large span of metallicities and temperature ranges within each spectral type. As such, our final distance estimates should be viewed as a first-order estimate. We find that F stars will likely saturate out to $\sim$ 1 kpc, G stars to $\sim600$ pc, and K stars to $\sim 400$ pc.

\begin{deluxetable*}{llcc}[h!]
\label{tab:GCNSLSST}
\tablecaption{Astrometry and Photometry of Gaia Catalog of Nearby Stars and LSST DP1 Cross-matches}
\tablehead{\colhead{Column Label} & \colhead{Description} & \colhead{Example} &\colhead{Units}}
\startdata
    GaiaEDR3 & Gaia EDR3 identifier & 4830033050425668608 & \dots \\
    objectId & LSST DP1 identifier (LSST-DP1-O-) & 592914119179382261 & \dots \\
    RA & LSST RA if unsaturated, Gaia EDR3 RA if saturated & $58.40687550$ & deg \\
    RA\_Err & Uncertainty of RA & $0.00000019$ & deg \\
    DEC & LSST DEC if unsaturated, Gaia EDR3 DEC if saturated  & $-48.78568358$ & deg \\
    DEC\_Err & Uncertainty of DEC& $0.00000029$ & deg \\
    Plx & Gaia EDR3 parallax & $9.76$ & mas \\
    Plx\_Err & Uncertainty of Gaia EDR3 parallax & $0.30$ & mas \\
    PMRA & Gaia EDR3 RA proper motion ($\mu_\alpha$) & $56.40$ & mas/yr \\
    PMRA\_Err & Uncertainty of Gaia EDR3 RA proper motion ($\mu_\alpha$) & $0.31$ & mas/yr \\
    PMDEC & Gaia EDR3 DEC Proper Motion ($\mu_\delta$) & $0.51$ & mas/yr \\
    PMDEC\_Err & Uncertainty of Gaia EDR3 DEC proper motion ($\mu_\delta$) & $0.36$ & mas/yr \\
    Gmag  & Gaia $G$ magnitude & 19.576 & mag \\
    Gmag\_Err & Uncertainty of Gaia $G$ magnitude & 0.005 & mag \\
    BPmag & Gaia $BP$ magnitude & 21.143 & mag \\
    BPmag\_Err & Uncertainty of Gaia $BP$ magnitude & 0.178 & mag \\
    RPmag & Gaia $RP$ magnitude & 18.066 & mag \\
    RPmag\_Err & Uncertainty of Gaia $RP$ magnitude & 0.024 & mag \\
    rmag & LSST $r$ magnitude & 21.415 & mag \\
    rmag\_Err & Uncertainty of LSST $r$ magnitude & 0.005 & mag \\
    imag & LSST $i$ magnitude & 18.917 & mag \\
    imag\_Err & Uncertainty of LSST $i$ magnitude & 0.002 & mag \\
    zmag & LSST $z$ magnitude & 17.74 & mag \\
    zmag\_Err & Uncertainty of LSST $z$ magnitude & 0.001 & mag \\
    ymag & LSST $y$ magnitude & 17.041 & mag \\
    ymag\_Err & Uncertainty of LSST $y$ magnitude & 0.001 & mag \\
    Saturated & Is the GCNS object saturated in LSST? & No & \dots \\
    SPT & Spectral type estimate (from this work) & M & \dots \\
\enddata
\tablecomments{The complete table is available in a machine-readable format in the online journal.}
\end{deluxetable*}

\section{UCD Candidates from DP1} \label{sec:candidateselection}
We search for brown dwarf candidates in the two DP1 fields with Euclid overlap using traditional color-color selection cuts. In the following subsections we describe our selection criteria and the resulting UCD candidates.

\begin{figure}[h!]
    \centering
    \includegraphics[width=\columnwidth]{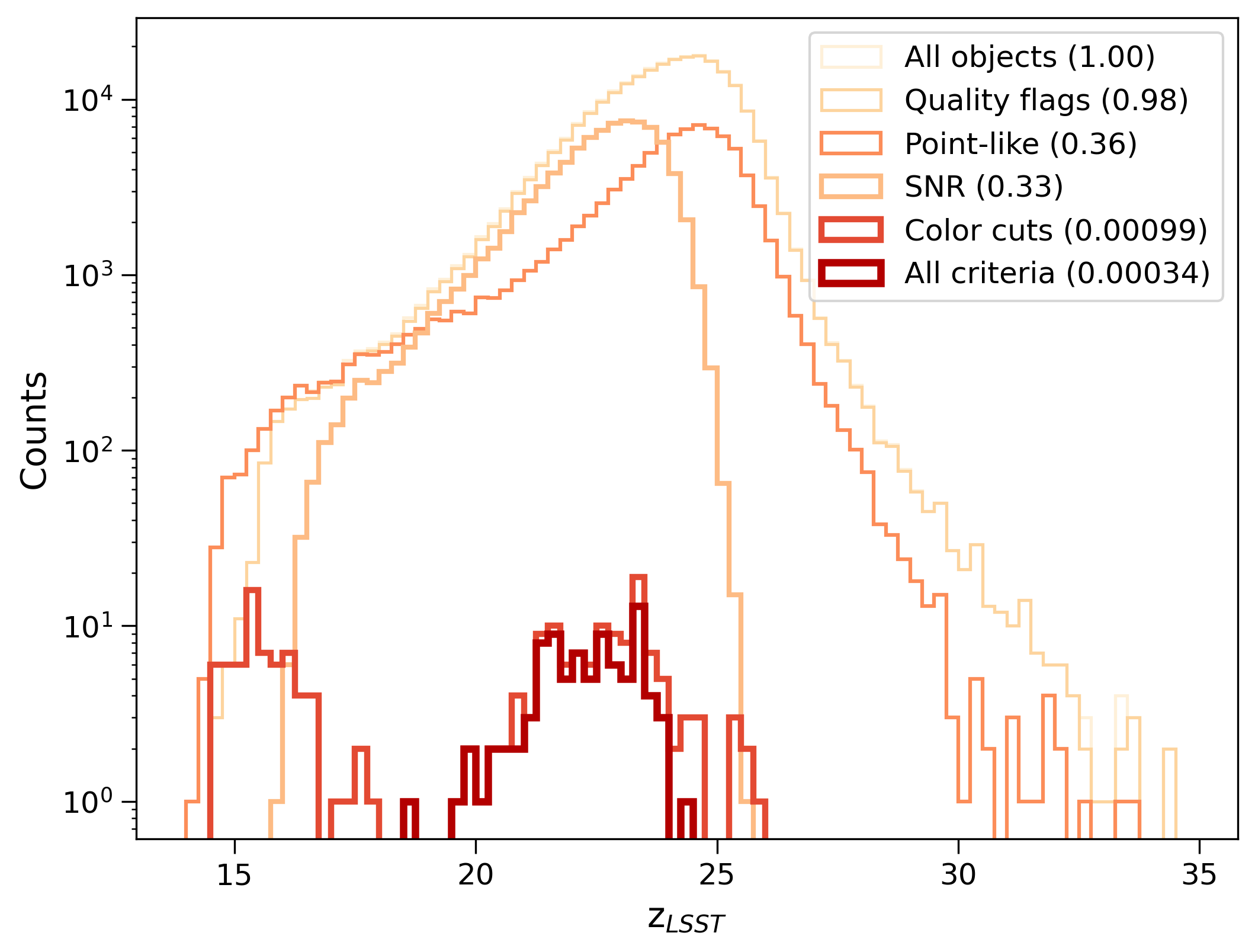}
    \caption{Number of objects as a function of LSST $z$ band magnitude that pass subsequent selection criteria. Starting from the 260,311 LSST and Euclid cross-matched objects, each selection criteria filter is shown with the remaining fraction of objects shown in the legend. The combination of all criteria retains 89 objects. The strictest criteria is the color-color selection region based off of Euclid-confirmed UCDs in \citet{Zerjal} and recovered DES UCD candidates from \citet{dalponte23}.}
    \label{fig:selectionhistogram}
\end{figure}

\subsection{LSST DP1 Data and Filtering}\label{sec:dataselection}
Of the seven DP1 fields, two overlap with the Euclid Quick Data Release (Q1): ECDFS and EDFS. For each of these fields, we query the DP1 Object table through the RSP for objects within 1.25$\degree$ from the field centers. We retrieved columns related to PSF photometry, position, extendedness, and pixel flags. The LSST DP1 Object table and Euclid Q1 MER Catalog were cross-matched by LSDB\footnote{\url{https://lsdb.io/}} \citep{lsdbBASE,lsdbDP1}, linking LSST's \texttt{objectId} to Euclid's \texttt{object\_id}. We retain objects with Euclid counterparts and query the Euclid MER catalog via TAP to retrieve object positions, detection quality flags, fluxes, and morphology measurements. Specifically, we query for the \texttt{flux\_*\_2fwhm\_aper} fluxes as recommended in \citet{Mohandansan25} and calculate AB magnitudes as described in the MER photometry cookbook\footnote{\url{https://euclid.esac.esa.int/dr/q1/dpdd/merdpd/merphotometrycookbook.html}} for Euclid's visible ($I_E$) and near infrared ($Y_E$, $J_E$, $H_E$) bands. To avoid confusion, Euclid bandpasses will be denoted using uppercase letters with the subscript $E$ ($I_E,Y_E, J_E, H_E$) while LSST bandpasses will be written in lowercase ($u, g, r, i, z, y$).

Across the ECDFS and EDFS fields, there are 260,311 LSST objects with a paired Euclid counterpart. Both fields have similar extinction values and no reason to suggest different selection criteria by field would be necessary. To arrive at our final selection of 89 UCD candidates (44 in ECDFS, 45 in EDFS), we apply a series of signal-to-noise ratio (SNR), quality, and color cuts, each explained in the follow subsections. The impact of each of these selection cuts on the dataset is visualized in Figure \ref{fig:selectionhistogram}.  

\subsubsection{Detection Quality Criteria}\label{sec:quality_flags}
The first selection criteria used LSST photometric quality flags to exclude objects with saturated pixels, detections near image edges, or general failures with PSF measurements. Specifically, we removed any objects with \texttt{*\_pixelFlag\_saturated}, \texttt{*\_pixelFlag\_saturatedCenter}, \texttt{*\_psfFlux\_flag\_edge}, or \texttt{*\_psfFlux\_flag} set to True in any of the $i,z$, or $y$ bands. The saturation flags allow us to remove any objects that would later be selected by color-color cuts but are actually artifacts created in the deblending process for saturated stars as discussed in Section \ref{sec:GCNSmatches}. We do not include the \texttt{detect\_isIsolated} flag in our detection quality criteria, as done in \citet{Zhang25}, since many potential candidates are near other objects and are listed as products of the deblender even without overlapping pixels. Using the \texttt{detect\_isIsolated} flag would remove $>90\%$ of objects that would meet all of our other selection criteria, including many potential UCDs, simply by being deblended. The detection quality flag criteria removes 5,564 of the original 260,311 objects (98\% remaining).

\subsubsection{SNR Criteria}\label{sec:snrcuts}
We then implement signal-to-noise ratio (SNR) requirements in $i,z,y$ to only retain well-measured objects that allow for the construction of optical colors. Since the $z$ band is the deepest band of the three in DP1, we required $\text{SNR}_{z}\geq5$ and allowed a less strict $\text{SNR}_{i,y}\geq 3$ in the $i,y$ bands. We considered requiring $\text{SNR}_r\geq3$, but ultimately found imposing restrictions on the $r$ band yielded candidate lists that were exclusively M dwarfs. Our analysis in this work centers around $i,z,y$ and we leave further investigation of the $g,r$ bands for later data products with LSSTCam instead of LSSTComCam. 

We also require that objects have $\text{SNR}_{I_E,Y_E,J_E,H_E}\geq3$. These requirements are more liberal than those implemented in \citet{Zerjal}, but the addition of optical SNR criteria still yields well-measured candidates. It is worth noting that the Euclid $I_E$ bandpass is a wide filter that spans wavelengths from the red cutoff of the LSST $g$ band to the blue cutoff of the LSST $y$ band. Utilizing both bands, despite overlap, and building both LSST- and Euclid-based optical colors allows for greater characterization of the optical portion of the UCDs' SEDs. We further explore colors with overlapping bandpasses in Section \ref{sec:colorcuts}. 

We validated our SNR criteria with infrared-selected UCD candidates from \citet{Zerjal}. 17 of their C1 class objects (candidates with confirmed UCD spectral features and matches to UCD standards) were matched to LSST counterparts by LSDB and all 17 significantly surpass $\mathrm{SNR}_{i,z,y}\geq5$. The joint optical and infrared SNR criteria remove 174,087 of the original 260,311 objects (33\% remaining). 84,929 (32\%) objects meet both SNR and detection quality requirements.

\begin{figure}[h!]
    \centering
    \includegraphics[width=\columnwidth]{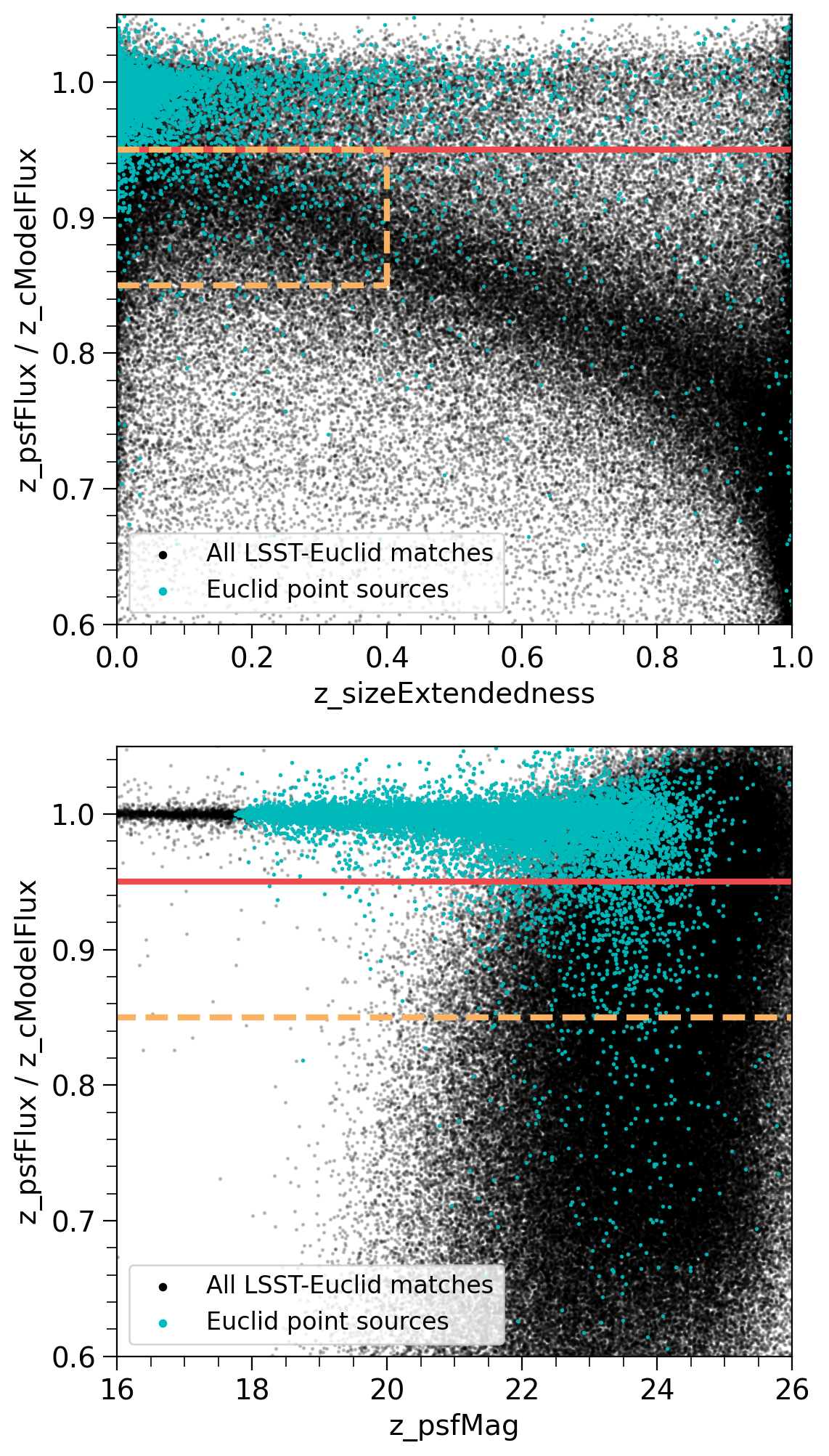}
    \caption{Extendedness in the LSST Object table parameters. Euclid objects meeting point-like criteria from \citet{Zerjal} are shown in magenta against all DP1 objects in  black. The upper panel shows how LSST's flux ratio (\texttt{z\_psfFlux}$/$ \texttt{z\_cModelFlux}, see Section \ref{sec:extendednesscuts} for details) \textit{vs} the \texttt{z\_sizeExtendedness} parameter. The lines represent the LSST pipeline cuts: above the solid red line (\texttt{z\_psfFlux}$/$ \texttt{z\_cModelFlux}$=0.95$), objects are classified as point-like in LSST. \texttt{z\_sizeExtendedness} is a moments-based approach where greater values indicate larger degrees of extendedness. As shown in the lower panel (\texttt{z\_psfFlux}$/$ \texttt{z\_cModelFlux} \textit{vs} \texttt{z\_psfMag}), a significant number of faint point sources ($z>22$) drop below the 0.95 flux ratio threshold. Our final extendedness criteria includes all objects with \texttt{z\_psfFlux}$/$ $\texttt{z\_cModelFlux}>0.95$ or $\texttt{z\_psfFlux}$/$ \texttt{z\_cModelFlux}>0.85$ and $\texttt{z\_sizeExtendedness}<0.4$ (dotted red lines).}
    \label{extendedness}
\end{figure}

\subsubsection{Extendedness Criteria}\label{sec:extendednesscuts}
Next, we remove any extended objects. The LSST Object table includes an extendedness flag for each band (\texttt{*\_extendedness}) and for the reference band (\texttt{*\_refExtendedness}). The extendedness parameter is a ratio of an object's PSF flux and CModel flux in a given band, where the CModel flux is a galaxy-oriented photometric measurement that linearly combines multiple PSF-convolved ellipses. The LSST science pipelines compute the ratio of PSF to Cmodel flux and assign an object to be point-like (\texttt{*\_refExtendedness}==0) if $\texttt{*\_psfFlux}/\texttt{*\_cModelFlux}\geq0.95$. However, for faint point sources at lower SNR, the CModel flux measurement can become unreliable and the extendedness parameter in different bands does not always agree. For UCD selection purposes, a more reliable star/galaxy separation metric is necessary.

To design a more robust approach for faint UCDs, we investigated how objects that meet Euclid point source requirements in \citet{Zerjal} were classified by the LSST pipelines. To briefly summarize, \citet{Zerjal} filtered point sources by requiring the following combination of morphology cuts: \texttt{ELLIPTICITY} $< 0.2$; \texttt{FWHM} $<1.5''$; \texttt{DET\_QUALITY\_FLAG} $\leq 3$; $-3.25 \text{ mag arcsec}^{-2} \leq \texttt{MUMAX\_MINUS\_MAG} \leq -2.65 \text{ mag arcsec}^{-2}$; and $\texttt{SNR$_{I_E, Y_E, H_E}$}>4$. In Figure \ref{extendedness} we show the flux ratio and moment-based extendedness values (\texttt{*\_sizeExtendedness}) for all objects with LSST and Euclid cross-matches and in magenta overlay Euclid-defined point sources. Since the $z$ band is deeper than the $y$ band in DP1, we use the $z$ band extendedness information to classify objects as extended or point-like. The primary extended object locus can be seen at lower flux ratios ($\sim 0.75$) and higher \texttt{z\_sizeExtendedness} values whereas the point source locus is centered at high flux ratio values near and above the pipelines' 0.95 cutoff. Some point sources extend down to a flux ratio of 0.85 and out to \texttt{z\_sizeExtendedness} $\sim$ 0.4 (dotted red lines in the upper panel of Figure \ref{extendedness}). The lower panel of Figure \ref{extendedness} shows that the vast majority of such objects are found at fainter magnitudes ($z\gtrsim22$) and, subsequently, lower SNR. For the purposes of discovering the faintest UCDs, we conclude that a more involved approach to star/galaxy separation is in order; anchored using the $z$ band, we retain objects that either meet the pipelines' 0.95 flux ratio cutoff or have both a $\texttt{*\_psfFlux}/\texttt{*\_cModelFlux}>0.85$ (lower dotted red lines in Figure \ref{extendedness}) and \texttt{z\_sizeExtendedness} $<0.4$. The combination of flux and moment-based approaches allows for a more dynamic filtering of faint point sources while avoiding the majority of extended extragalactic objects. The extendedness criteria removes 167,529 of the original 260,311 objects (36\% remaining). Combining the detection quality, SNR, and extendedness criteria retains 26,146 (10\%) of the original objects. 

\subsubsection{Color-Color Criteria}\label{sec:colorcuts}
The color-color space occupied by UCDs in LSST is currently largely unverified. While predictions can be made from the similar Pan-STARRS photometric system, the precise color terms between the two types of LSST CCDs or atmospheric water features may lead to significant, unexpected differences between the two systems, particularly in the $y$ band. The best practice is to use spectroscopically confirmed objects to identify the locus of UCDs in a specific photometric system. As described above, however, at this time, an empirical sample of known and confirmed UCDs within the ECDFS and EDFS footprints is limited. Thus, using the spectroscopically confirmed sample of UCDs from \citet{Zerjal}, we aim to find the most effective combination of LSST and Euclid filters to distinguish UCDs from similarly colored background objects, namely extragalactic sources. 

For all cross-matched objects, we construct colors using $i,z,y,I_E,Y_E,J_E,H_E$. We use both Euclid's visible band, $I_E$, as well as LSST's three optical bands since the $I_E$ band is a much broader bandpass better suited to measuring candidates' overall optical flux while LSST's narrower filters can probe specific spectral features over the same wavelengths more precisely (e.g. Na I at 589 nm in $i$ or K I at 768 nm in $z$). Both the Na I and K I features have been historically difficult to reproduce in models \citep{allard03,allard05,allard07KH,allard12}, and including Euclid's $I_E$ and LSST's $i,z$ is important to derive color selection criteria that robustly covers the region until larger empirical samples of objects are available.  
We also construct colors using both $y$ and $Y_E$ since they have slightly different bandpasses but significantly different PSFs ($\sim0.6\arcsecond$ for Euclid and $\sim1.2\arcsecond$ for LSST). Furthermore, since Euclid is space-based, we are not concerned with atmospheric transmission effects that may plague the $y$ band and require additional study in future LSST datasets with LSST's survey camera, LSSTCam.

We construct a total of 19 colors and determine the range occupied by the C1 and C2 class candidates in \citet{Zerjal}. The C2 class candidates are objects with clear UCD-like features but no close matches to any template spectra. We include the C2 class candidates now to assemble a broader sample of objects which may include color outliers, chemically peculiar objects, young objects, low-metallicity subdwarfs, or unresolved binaries.
From the 106 C1 and C2 template objects that are within the ECDFS and EDFS fields, we determine the upper and lower bounds in the color space occupied by the C1 and C2 class objects. For optical colors, we also include the 10 DES cross-matched objects discussed in Section \ref{sec:DESmatches}. We further extend each color range to the nearest tenth of a magnitude to allow for additional scatter (taking the floor for lower limits and ceiling for upper limits). The final colors and ranges used are shown in Table \ref{colorcutvalues}. While 19 colors is certainly a large number of criteria and redundant, by constructing colors with overlapping bandpasses ($i,z,y$ vs $I_E$ and $y$ vs $Y_E$) we can build our understanding of LSST colors to reduce the number of constraints in the future.

\begin{table}[ht]
\centering
\caption{Color Selection Criteria}
\label{colorcutvalues}
\begin{tabular}{l c c} 
\hline
\hline
Color & Lower Limit & Upper Limit\\
 & (mag) & (mag)\\
\hline
&\textbf{LSST only}& \\
\hline
$i-z$     & $0.9$ & $3.2$ \\
$i-y$     & $1.7$ & $4.1$ \\
$z-y$     & $0.4$ & $1.5$ \\
\hline
&\textbf{Euclid only}& \\
\hline
$I_E-Y_E$ & $2.5$ & $3.5$ \\
$I_E-J_E$ & $0.6$ & $3.8$ \\
$I_E-H_E$ & $2.7$ & $4.2$ \\
$Y_E-J_E$ & $-2.0$& $0.6$ \\
$Y_E-H_E$ & $0.1$ & $1.0$ \\
$J_E-H_E$ & $0.0$ & $2.2$ \\
\hline
&\textbf{LSST \& Euclid}\\
\hline
$i-Y_E$   & $1.6$ & $4.5$ \\
$i-J_E$   & $0.5$ & $4.8$ \\
$i-H_E$   & $2.0$ & $5.2$ \\
$z-Y_E$   & $0.4$ & $1.4$ \\
$z-J_E$   & $-0.9$& $1.9$ \\
$z-H_E$   & $0.8$ & $2.3$ \\
$I_E-y$   & $2.2$ & $3.1$ \\
$y-Y_E$   & $-0.5$& $0.5$ \\
$y-J_E$   & $-0.4$& $1.0$ \\
$y-H_E$   & $-0.4$& $1.3$ \\
\hline
\hline

\end{tabular}
\end{table}

From examining the color space occupied by candidates, we can determine which colors are most and least efficient at UCD selection. For example, as seen in Table \ref{colorcutvalues}, we select UCDs from $-0.9 \leq (z-J_E) \leq 1.9$. However, $95\%$ of LSST and Euclid cross-matches fall within $-0.95\leq(z-J_E)\leq2.1$, making $z-J_E$ a weaker choice for color selection processes unless coupled with additional colors. Following this logic, the best colors, where the candidate selection region is most distinct from the primary stellar locus, are $i-z$, $i-y$, $I_E-Y_E$, $I_E-H_E$, $i-Y_E$, $i-H_E$, and $I_E-y$. Notably, the colors that best distinguish UCDs include either $i$ or $I_E$ while colors involving $z,y$ with later bands do not significantly accentuate UCDs beyond the 95\% distributions of other objects. This has significant implications: UCDs are intrinsically faint in optical wavelengths and the faintest UCDs are likely to drop out of optical detectability ranges, complicating their classification.

After applying our color criteria, 254 of the original 260,311 objects (0.1\%) remain. Combining all four selection criteria (detection quality, SNR, extendedness, and color), further reduces the sample until we reach our final set of 89 candidates (0.03\% of the original sample) across the two fields. The progressive filtering done is shown in Figure \ref{fig:selectionhistogram}. After applying our selection criteria, we visually inspected all 89 candidates in both LSST and Euclid images to make sure no artifacts or clearly extended objects remained. No such issues were present with the 89 candidates and all selections were retained.

\subsection{Selected Candidates}\label{sec:candidates}
\begin{figure*}[ht!]
    \centering
    \includegraphics[width=\textwidth]{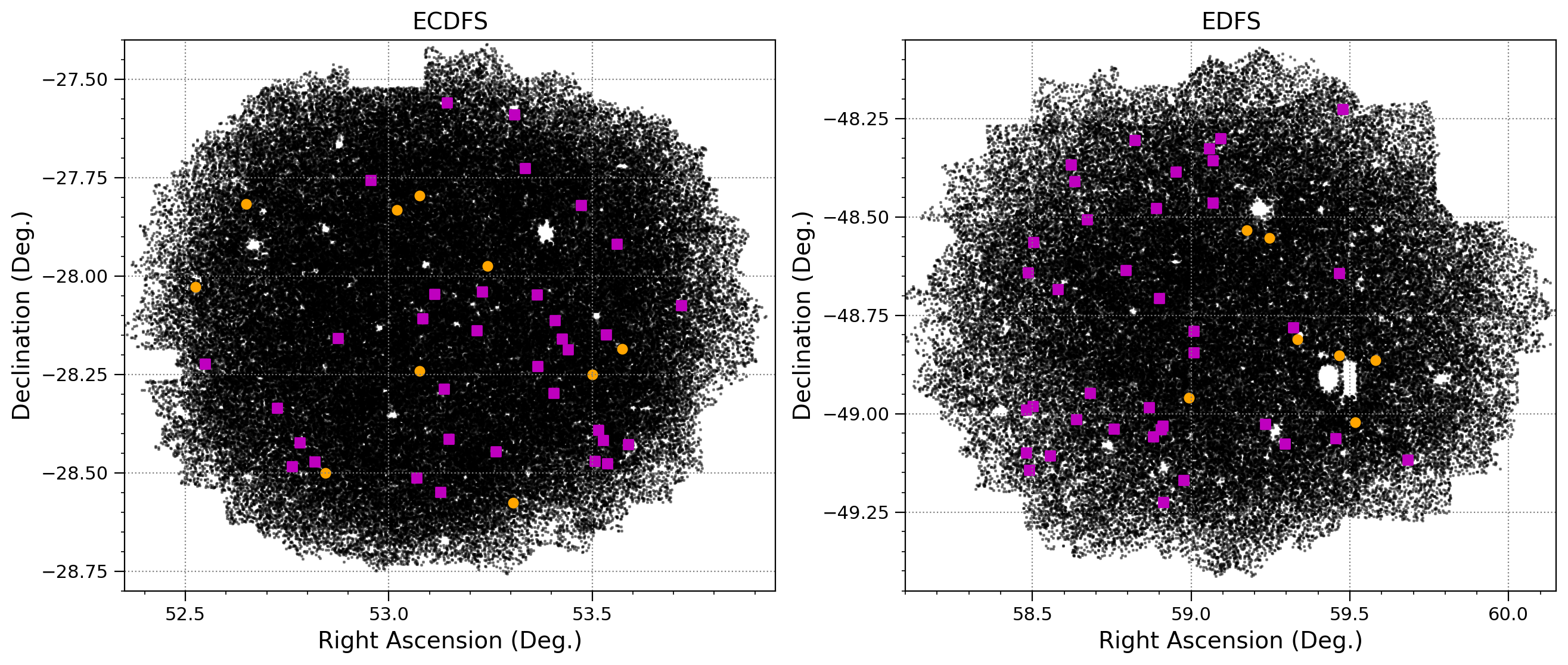}
    \caption{Candidate object positions in both the ECDFS (left) and EDFS (right) fields are overplotted on all objects with LSST and Euclid cross-matches, shown as black dots. Of the 89 candidates, 45 are in ECDFS and 44 are in EDFS. Orange dots represent candidates unique to this work, purple squares are candidates also found in \citet{Zerjal}, \citet{Zhang25}, or \citet{dalponte23}.}
    \label{candpositions}
\end{figure*}

\begin{figure}[h!]
    \centering
    \includegraphics[width=\columnwidth]{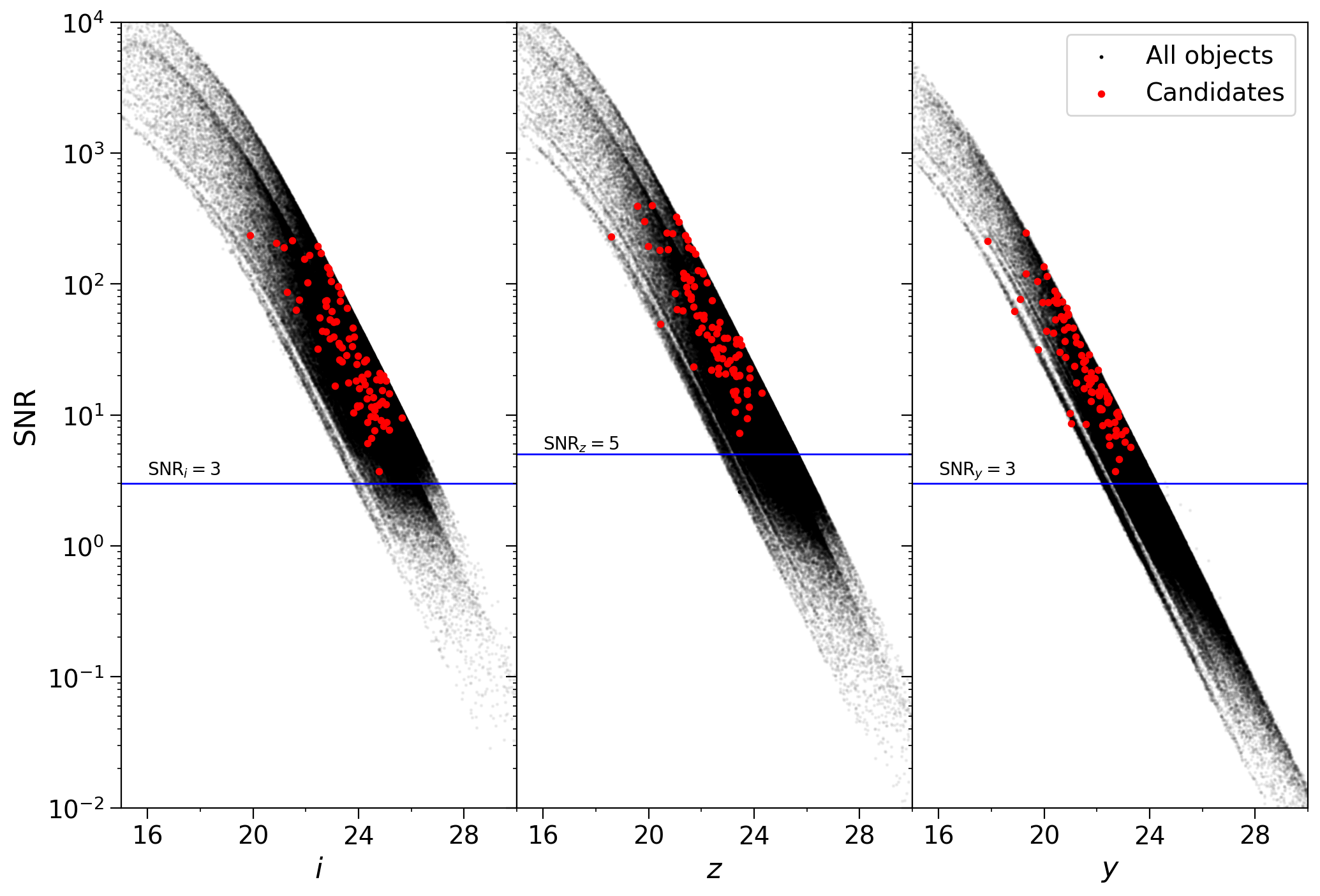}
    \caption{Final candidate objects' SNR in the LSST $i,z,y$ bands are shown in red dots against all objects with LSST and Euclid cross-matches. $\mathrm{SNR}$ decreases as object magnitudes increase. Applying more conservative SNR cuts in the $z$ band compared to the $i,y$ bands allow for the selection of fainter candidates.}
    \label{fig:candidatesnr}
\end{figure}

We present the 89 selected candidates in Table \ref{tab:candidates}. In Figure \ref{candpositions} we show the positions of the 44 candidates within the ECDFS field and the remaining 45 in the EDFS field. A majority of the candidates have been previously discovered by \citet{Zerjal}, \citet{Zhang25}, or \citet{dalponte23}. In ECDFS, we found 10 new candidates, confirmed 28 from \citet{Zerjal}, 1 from both \citet{Zerjal} and \citet{Zhang25}, and 5 from \citet{Zerjal} and \citet{dalponte23}.  Similarly, in EDFS, we present 7 new candidates, recover 33 from \citet{Zerjal}, confirmed 4 from both \citet{Zerjal} and \citet{dalponte23}, and 1 from \citet{Zerjal}, \citet{Zhang25}, and \citet{dalponte23}. We denote new and previously discovered objects with the ``CRef'' column in Table \ref{tab:candidates}. There are additional objects within both fields discovered by \citet{Zhang25} and \citet{Zerjal} that do not meet our SNR or extendedness criteria and thus are excluded from our candidate list. For example, some objects in \citet{Zhang25} are only detected in the $y$ band and are excluded in our criteria since we cannot construct optical colors for such objects. In Figure \ref{fig:candidatesnr} we show our candidates' $\mathrm{SNR}_{i,z,y}$ compared to our selection criteria. Retained candidates reach $i=25.65$ mag, $z=24.30$ mag, and $y=23.29$ mag, significantly fainter than the majority of known UCDs with optical photometry.

\begin{figure*}[t!]
    \centering
    \includegraphics[width=\textwidth]{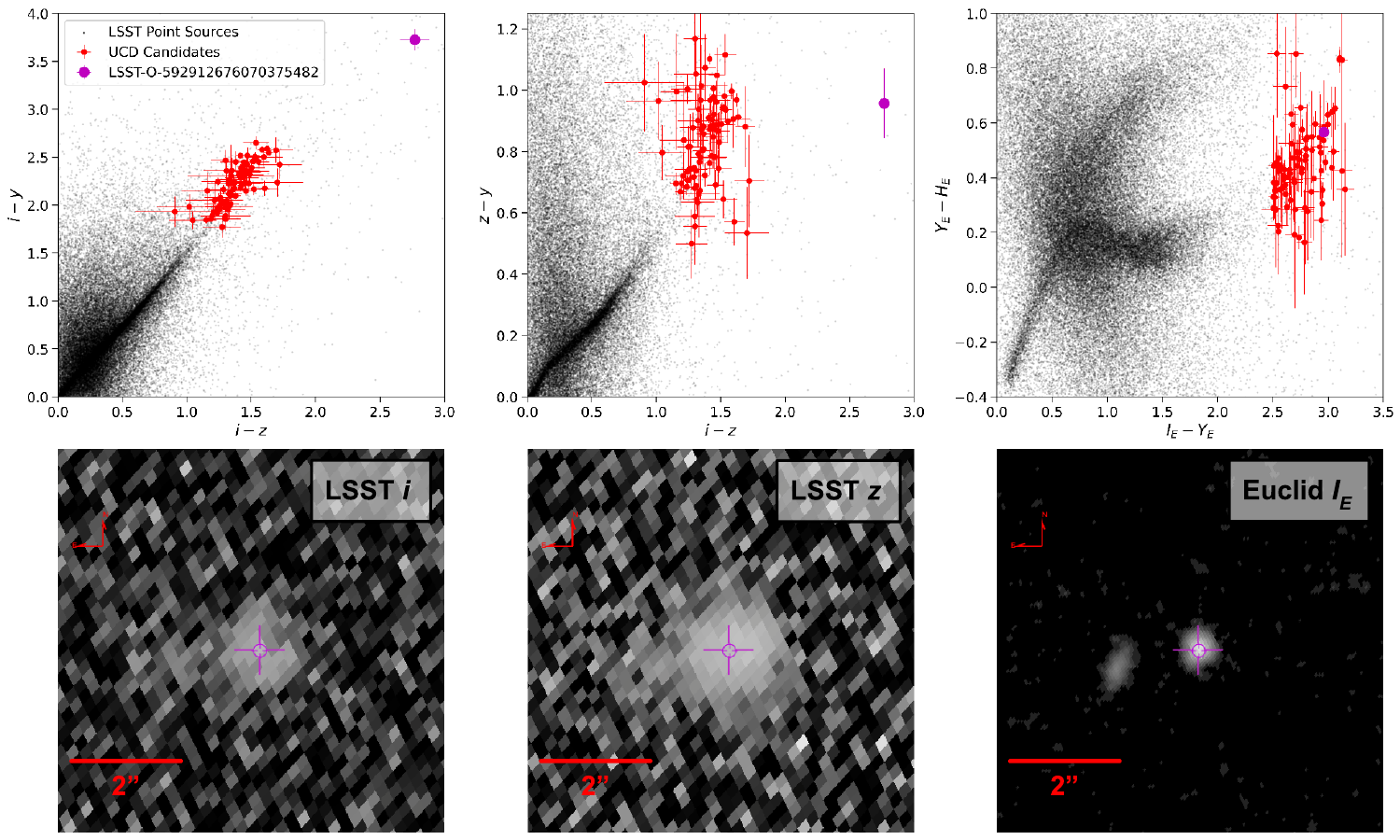}
    \caption{Candidates are shown in LSST-only color-color plots (upper left and upper middle panels) as well as a Euclid-only color-color plot (upper right panel). Candidates are shown as red points with their color uncertainties while LSST point sources detected in both LSST and Euclid are plotted as black dots. One candidate, LSST-O-592912676070375482 (hereafter J0354-4900), is marked in magenta across all three upper panels. J0354-4900 is an optically red outlier in $i$-based colors but does not stand out from other candidates in $I_E$-based colors. On the bottom row of panels, we show the LSST $i$ (lower left), $z$ (lower middle), and the Euclid $I_E$ (lower right) coadds for J0354-4900, marked by the magenta crosshair. All three cutout images are 6.9\arcsecond$\times$6.9\arcsecond. In the LSST images, a nearby ($\sim 1.5\arcsecond$ separation) galaxy is blended with the candidate. However, with Euclid's space-based PSF, the two objects can be clearly separated into a galaxy and point source.}
    \label{fig:opticalcolors}
\end{figure*}

\begin{deluxetable*}{llcc}\label{tab:candidates}
\tablecaption{UCD candidates in LSST DP1's ECDFS and EDFS fields}
\tablehead{\colhead{Column Label} & \colhead{Description} & \colhead{Example} &\colhead{Units}}
\startdata
objectId& LSST DP1 identifier (LSST-DP1-O-) &$592913088387238886$ & \dots \\
object\_id& Euclid QR1 identifier &$-595810742488645292$ & \dots \\
RA & LSST RA &$59.58108532$ & deg \\
RA\_Err & Uncertainty of LSST RA & 0.00000959 &  deg \\
DEC & LSST DEC & $-48.86452976$& deg \\
DEC\_Err & Uncertainty of LSST DEC & 0.00001166 & deg \\
imag & LSST $i$ magnitude & $24.233$ & mag \\
imag\_Err & Uncertainty of LSST $i$ magnitude & $0.064$ & mag \\
zmag & LSST $z$ magnitude &$22.677$ & mag \\
zmag\_Err & Uncertainty of LSST $z$ magnitude &$0.033$ & mag \\
ymag & LSST $y$ magnitude &$21.779$ & mag \\
ymag\_Err & Uncertainty of LSST $y$ magnitude & $0.052$& mag \\
IEmag & Euclid visible $I_E$  magnitude & $24.246$ & mag \\
IEmag\_Err& Uncertainty of Euclid visible $I_E$ magnitude &$0.066$ & mag\\
YEmag & Euclid $Y_E$ magnitude &$21.569$ & mag \\
YEmag\_Err & Uncertainty of Euclid $Y_E$ magnitude & $0.037$& mag \\
JEmag & Euclid $J_E$ magnitude & $21.180$& mag \\
JEmag\_Err & Uncertainty of Euclid $J_E$ magnitude & $0.024$& mag \\
HEmag & Euclid $H_E$ magnitude &$20.975$ & mag \\
HEmag\_Err&  Uncertainty of Euclid $H_E$ magnitude & $0.021$& mag \\
Phototemp & Photometric temperature estimate from this work & $2400$ & K \\CRef\tablenotemark{a} & Cross-referenced literature containing this object & 0 & \dots \\
\enddata
\tablecomments{The complete table is available in a machine-readable format in the online journal.}
\tablenotetext{a}{Cross-references are coded such that: 0 - Unique to this work; 1 - Also in \citet{Zerjal}; 2 - Also in \citet{Zhang25}; 3 - Also in \citet{dalponte23}.}
\end{deluxetable*}

We show the candidates in two LSST-only and one Euclid-only color-color plot in Figure \ref{fig:opticalcolors}. The majority of candidates form an extension to the arm of the stellar locus. The one extreme red outlier in optical plots, LSST-O-592912676070375482 (hereafter J0354-4900), is marked in magenta and sits within the main grouping of candidates in the Euclid color-color plot. J0354-4900 is underluminous in $i$ compared to other candidates but normal in $z,y$. Photometric uncertainties for the object's $i$ band are larger ($\sim0.1$ mag) than for the $z,y$ bands ($\sim0.01$ mag). J0354-4900 was one of the DES counterparts discussed in Section \ref{sec:DESmatches} and is not a red outlier in DES photometry. Visual inspection of both LSST and Euclid images reveal a nearby galaxy (LSST-O-592912676070375483, $\sim$1.5\arcsecond separation) that is well separated from J0354-4900 in Euclid's $I_E$ band but is moderately blended in all other bands except for $y$. The galaxy is not detected in $y$. The Euclid $I_E$ band also clearly shows J0354-4900 is a distinct point source with a nearby galactic neighbor and not an extended object. The added value from space-based imaging with sharp PSFs cannot be overstated for star/galaxy separation of faint objects. In the LSST pipelines, the galaxy and UCD candidate share the same \texttt{parentObjectId}, meaning they were deblended from the same footprint.  We retrieved the Euclid spectrum for J0354-4900 and compared it to UCD spectral standards using SPLAT \citep{splat}. We found J0354-4900 was best matched by the L1 standard 2MASSW J2130446-084520 \citep{Bardalez14}. The candidate shows strong H$_2$O absorption features that support an L dwarf classification. Despite its exceptionally red optical appearance, there were no youth signatures in the spectral $H$ band region. 

Given the large number of faint galaxies LSST and Euclid expect to discover, deblended photometry in cases like this will be a point of concern that users must account for in future data releases. Larger samples of objects with nearby neighbors will be critical to understanding best practices with deblended objects. 

\begin{deluxetable*}{cccc}\label{tab:newobjects}
\tablecaption{New UCD candidates unique to this work}
\tablehead{\colhead{LSST ID} & \colhead{Euclid ID} & \colhead{RA\tablenotemark{a}} &\colhead{Dec\tablenotemark{a}} \\ \colhead{LSST-DP1-O-} & \colhead{} & \colhead{hh:mm:ss} & \colhead{dd:mm:ss}}
\startdata
592912469911949248 & -595159828490212530 & 03:58:03.83 & -49:01:16.5 \\
592913088387238886 & -595810742488645292 & 03:58:19.46 & -48:51:52.3 \\
592913157106711625 & -594657596488527189 & 03:57:51.78 & -48:51:09.9 \\
592913294545663770 & -589926114489600687 & 03:55:58.22 & -48:57:36.2 \\
592913844301475139 & -593355476488116819 & 03:57:20.53 & -48:48:42.1 \\
592914600215726820 & -592461704485544037 & 03:56:59.07 & -48:33:15.9 \\
592914600215728764 & -591754925485342024 & 03:56:42.12 & -48:32:03.1 \\
609781520902663878 & -528443113285004828 & 03:31:22.63 & -28:30:01.7 \\
609788117972430474 & -533063258285767848 & 03:33:13.52 & -28:34:36.5 \\
611253560813816607 & -535002664282505237 & 03:34:00.07 & -28:15:01.9 \\
611253766972261787 & -530762753282415330 & 03:32:18.30 & -28:14:29.6 \\
611254248008592910 & -535740731281853027 & 03:34:17.78 & -28:11:07.0 \\
611254591605987636 & -525246558280277292 & 03:30:05.92 & -28:01:39.9 \\
611255072642314468 & -532423102279750234 & 03:32:58.15 & -27:58:30.1 \\
611255828556559519 & -530196424278331491 & 03:32:04.71 & -27:49:59.3 \\
611255828556562234 & -530758584277963269 & 03:32:18.20 & -27:47:46.8 \\
611255965995500477 & -526492580278177462 & 03:30:35.82 & -27:49:03.9 \\
\enddata
\tablenotetext{a}{From LSST}
\end{deluxetable*}
\subsubsection{Phototemp Estimate}\label{sec:phototemp}
To characterize our candidates, we showcase the potential of a phototype algorithm as originally described in \citet{skrzypek15}. Phototype algorithms classify an object by comparing the object's photometry to a series of templates. This method is extremely powerful for the classification of objects in regions without spectroscopic follow up and well suited to large surveys. Ideally, the templates should be empirical samples of objects in the photometric systems available, but such a sample is not yet possible. We show the potential of a phototype method by combining LSST photometry with Euclid, but the concept can easily be extended to other infrared surveys (e.g. VHS, Roman, CatWISE). 

Since we do not have a readily available sample of well understood objects in the LSST photometric system, we instead build our phototype estimator using SANDee, a grid of model atmospheres \citep{sandee}. The SANDee models span from $0.0001 - 999.5$ $\mu$m and describe objects with temperatures from $T_{\text{eff}} = 700$ K to $4,000$ K, metallicities from [Fe/H]$ = -2.4$ to $+0.3$, surface gravities from $\log{g} = 2$ to $6$, and alpha-enhancement from [$\alpha$/H]$ = -0.05$ to $ 0.4$.  The SANDee models are an extension to the SAND evolutionary models \citep{sand}, both of which are specifically designed for low-mass stars and brown dwarfs. We compute synthetic colors in the LSST and Euclid bandpasses for all SANDee models. The LSST filter transmission profiles were retrieved from the LSST project's GitHub repository\footnote{\url{https://github.com/lsst-pst/syseng_throughputs/tree/main}} and the Euclid bandpasses were retrieved from the Spanish Virtual Observatory (SVO; \citealt{SVO1,SVO2,SVO3}).

For our phototype estimator, we only consider SANDee models with alpha-enhancement $-0.05\leq[\alpha\mathrm{/H}]\leq+0.2$, metallicities $-1.1\leq[\mathrm{Fe/H}]\leq+0.3$  and surface gravities  $3.5\leq\log{g}\leq5.0$. The alpha-enhancement and metallicity ranges were selected to best represent primarily galactic thin disk populations with some contributions from thick disk population properties \citep{Mackereth19}. We employ all the model temperatures to include negative samples for objects that are hotter than the late M and L dwarfs we seek to classify. Such objects should not be selected through our color criteria, but we include the templates to classify them should they pass all selection criteria. To create model templates, we compute the median color values for all models of a given temperature, resulting in a total of 30 templates ranging from $800$ K to $4,000$ K in steps of $100$ K. Next, we classify our candidates using a $\chi^2$ criterion, detailed in \autoref{phototemp_details}. This process assigns a  single best-fitting template for each candidate as the candidate's phototype. However, since our templates are labeled by temperature and not spectral type, we refer to the best-fitting template as the candidate's ``phototemp''.

\begin{figure*}[t!]
    \centering
    \includegraphics[width=\textwidth]{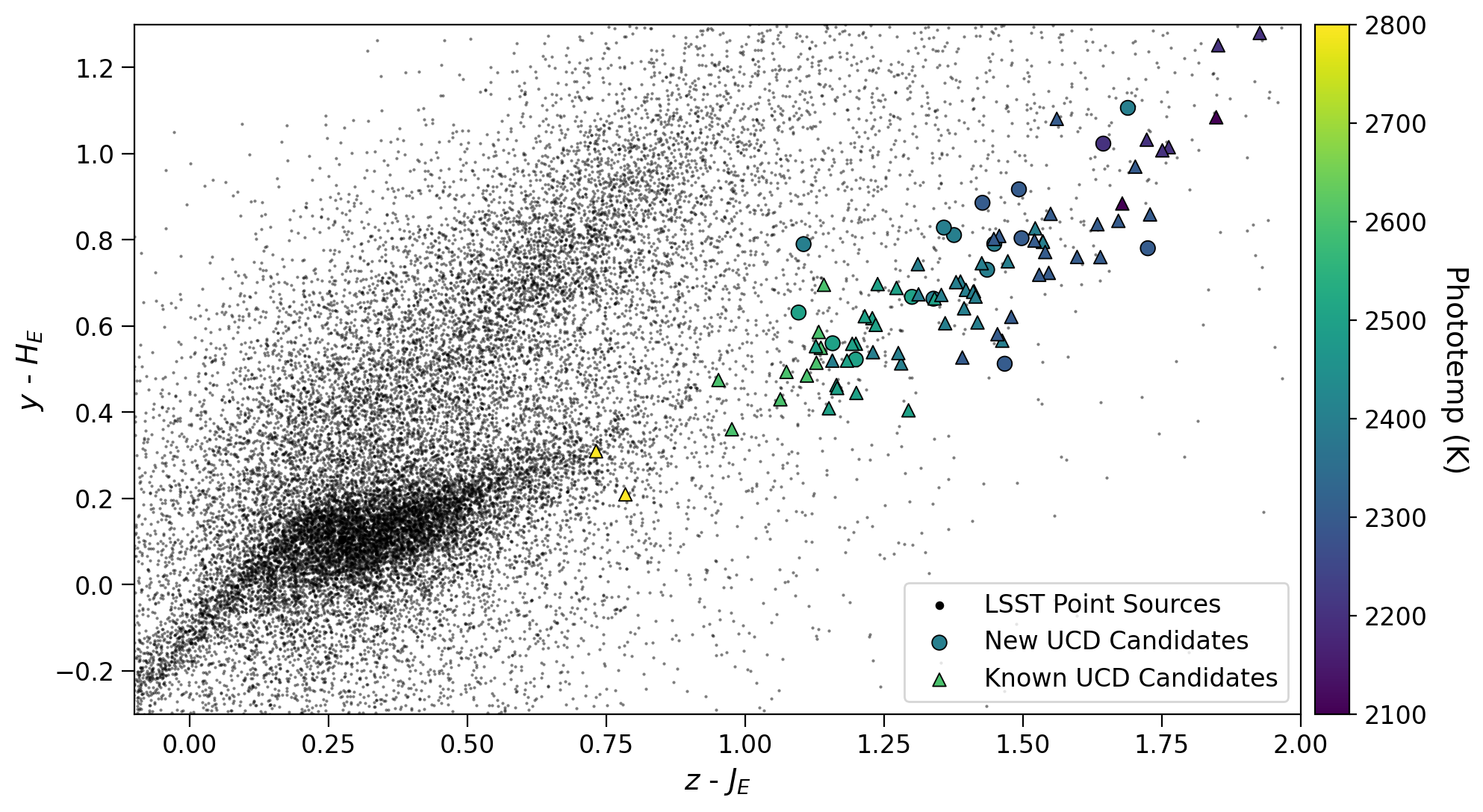}
    \caption{Candidate objects are shown with colored points against LSST point sources (black dots). The candidates are colored by their best matching template temperature, determined by phototype using the SANDee models. Candidate marker shapes are such that circles are unique to this work and triangles are objects also found in \citet{Zerjal}, \citet{Zhang25}, or \citet{dalponte23}. Combining LSST and Euclid filters allows for a clear temperature sequence to be seen in $z-J_E$ for late M and early L type dwarfs.}
    \label{phototemp}
\end{figure*} 
With each candidate having an assigned phototemp, we show the candidates using combined optical and NIR colors in Figure \ref{phototemp}. A clear temperature sequence can be seen as $z-J_E$ and $y-H_E$ increase. The warmest candidates have phototemps reaching 2,800 K, roughly equivalent to M7 spectral types, and the coolest candidate phototemps reach 2,100 K, entering the early L dwarf sequence \citep{Kirkpatrick21}. Of our 89 candidates, 31 have phototemps $\leq 2,300$ K, consistent with L dwarf temperatures (18 in ECDFS, 13 in EDFS). One of the two coldest object is indeed the extreme red outlier discussed in Section \ref{sec:candidates}, LSST-O-592912676070375482, with has a phototemp of 2,100 K, consistent with an early L1/L2 spectral type \cite{Kirkpatrick21}. This result agrees with the Euclid spectrum for this object.

In \citet{Zerjal}, $I_E-Y_E$ is used as a spectral type proxy, where objects with $2.5 \leq I_E-Y_E \leq 2.9$ are likely late-M dwarfs and objects with $2.9 \leq I_E-Y_E\leq 3.8$ are L dwarfs. Of our 89 candidates, 66 have $I_E-Y_E$ between 2.5 and 2.9. These 66 objects have phototemps centered around $2450\pm150$ K with a distinctly decreasing trend as $I_E-Y_E$ increases. The remaining 23 candidates have $I_E-Y_E$ colors between 2.9 and 3.2; all 23 of these candidates have phototemps centered around $2250\pm100$ K, most concentrated at 2,200 K and 2,300 K. The phototemps for these 23 candidates are comparable to objects with very late M or early L spectral types. To further validate our phototemps, we examined available Euclid spectra for our candidates. A substantial portion of these spectra showed M dwarf features and a few showed L dwarf features. However, the majority of the coolest phototemp candidates' spectra did not contain sufficient signal to definitively classify as UCDs.  None of the candidate spectra showed spectral signatures that would be expected of background extragalactic contaminants.

\section{Upcoming LSST DP2 Predictions}\label{future}
In this section, we preview upcoming LSST Data Preview 2 (DP2) fields and relative depths. In each case, we discuss predicted yields of UCDs (specifically brown dwarfs) and draw attention to key areas with potential difficulties.

\subsection{Road to LSST Data Release 1}
As of September 21, 2025, the Rubin Observatory completed Science Verification surveys and the formal 10-year survey operations is anticipated to begin in mid 2026. The survey timeline and scheduled data releases have fluid target dates, but the second Data Preview (DP2) is expected to be released in June - September 2026. Early science plans for DP2 and DR1 are discussed in great depth in \cite{RTN-011}. We briefly expand on two key components of DP2 and their implications for UCD science below.

\subsubsection{Deep Drilling Fields}

LSST's 10-year survey plan includes a Wide-Fast-Deep survey, covering 19.6k deg$^2$. Additionally, the survey will focus on six $\sim$ 1.8 deg$^2$ Deep Drilling Fields (DDFs) to reach deeper photometric depth and enhanced time-domain coverage. Observations of all six DDFs will be included in DP2 with varying completeness and photometric depth by field; exact visit counts and coadded depths can be found in \citet{RTN-011} and are reproduced in Table \ref{tab:dp2depths}. Two fields (ELAISS1, COSMOS) have visits in all six bands, three fields (EDFS\_A, EDFS\_B, ECDFS) have visits in the $g, r, i, z$ bands, and the final field (XMM\_LSS) was only observed in the $u$ band. While the DP2 depths for the DDFs will not reach full 10-year survey depths, ELAISS1 and COSMOS will exceed DP1 coadded depths while EDFS\_A, EDFS\_B, and ECDFS are expected to reach comparable depths to DP1 fields. For UCD science, the increased depth will allow for the discovery of some fainter (i.e. later spectral types), nearby objects but will primarily result in the detections of warmer objects at larger distances. Recovering large quantities of known brown dwarfs and discovering our coolest neighbors is best done with a wide field approach.

\subsubsection{Science Verification Surveys}
In addition to the six Deep Drilling Fields, LSSTCam commissioning included multiple wide area surveys collectively referred to as the Science Verification (SV) surveys. SV survey specifics are discussed in \citet{SITCOMTN-170} as well as on the Rubin Community Forums.\footnote{\url{https://community.lsst.org/}} The SV surveys were designed to test the LSST survey cadence, produce sky templates, and commission LSSTCam. While numerous small surveys were completed, the wide survey will be the most impactful SV component for UCDs. The SV wide area includes a 3,000 deg$^2$ region along the ecliptic plane, a 750 deg$^2$ area contained within, and a 300 deg$^2$ area contained within, each reaching deeper photometric depths. The 3k deg$^2$ component is estimated to reach 5$\sigma$ coadded depths of $(u,g,r,i,z,y) = (24.2,25.0,24.9,24.5,23.8,22.6$ mag). For comparison, the Dark Energy Survey Data Release 2 (DES DR2) covered 5,000 deg$^2$ to median depths of $(g,r,i,z,Y)=(24.7,24.4,23.8,23.1,21.7$ mag) at $\mathrm{SNR} \sim 10$ \citep{DESDR2}. The SV Wide Area survey will provide a wide field approach while still reaching new photometric depths, enabling the discovery of UCDs.

\subsection{Predicted Brown Dwarf Counts}\label{predictions}
To estimate potential brown dwarf yields in DP2, we employ the synthetic brown dwarf population created and described in depth in \citet{Honaker25}. To briefly summarize, the population models L0 - Y2 brown dwarfs in the Solar Neighborhood using recent \textit{Gaia}-derived star formation rate histories \citep{mazzi}, the substellar initial mass function \citep{Kirkpatrick21}, age-metallicity relation \citep{daltio_AMR_2021}, and local space densities \citep{best24}. The simulation extends vertically 1.3 kpc above and below the Galactic Plane as well as 200 pc radially outward from the Sun. For this work, we extend this simulation radially outward from 200 pc to 1 kpc. In doing so, we assume the brown dwarf spatial distribution and local star formation rate does not change significantly along the Galactic Plane in the additional volume, both reasonable assumptions since Galactic radial scale length is $\sim$ 2.6 kpc \citep{Juric08}. Since the scale height of the thick disk is 900 pc, the population is almost entirely comprised of thin and thick disk objects, with little to no contribution from halo populations. Using the star formation rate histories, local space densities, initial mass function, and age-metallicity relation, the simulation determines the number of objects in cylindrical slices of the simulated sphere and assigns each object an age, mass, and metallicity. Then, the simulation interpolates these parameters over the Sonora Diamondback evolutionary models \citep{diamondback} to assign objects temperatures, radii, and luminosities. We translate the temperatures into spectral types (L0-Y2) using temperature-spectral type relations from \citet{Kirkpatrick21}.

The original simulated population lacks photometry, but includes object spectral types and 3D positions. Using the Ultracool Sheet \citep{ucs}, we determine the median Pan-STARRS (PS1) absolute $z$ magnitude as a function of spectral type. Based on the absolute magnitude-spectral type relation, we assign simulated objects absolute $z$ magnitudes with a 1$\sigma$ scatter and  compute apparent $z$ magnitudes using their simulated positions. We also determined median $i-z$ and $z-y$ PS1 colors as a function of spectral type and assign apparent $i, y$ magnitudes to the simulated population with a 1$\sigma$ scatter. For spectral types later than T5, there were not enough objects with optical PS1 $i$ band photometry to determine a statistically significant color-spectral type relation. 

\citet{Zhang25} recently showed using atmospheric models and synthetic photometry that differences between the PS1 and LSST photometric systems for brown dwarfs are $\sim$0.02 mag for $i,z$ and $\sim$0.1 mag for $y$ ($y_{LSST} - y_{PS1} \sim - 0.1$).\footnote{In \citet{Zhang25}, the $y$ difference between the LSST and PS1 systems varies from 0.0 to -0.5 mag, but we are primarily concerned with L dwarfs (1400 K - 2400 K) where LSST will be most competitive. For L dwarfs, the photometric offset is centered at -0.1 mag.} Since 0.02 mag is smaller than the applied scatter for both the simulated $i$ and $z$ magnitudes, we simply adopt the PS1 $i,z$ values as LSST magnitudes. For the $y$ band, we subtract 0.1 mag from the simulated values to obtain LSST magnitudes. While a more rigorous approach is possible, our aim is to broadly estimate the number of brown dwarfs that may be detected in DP2 and the minor differences introduced between the PS1 and LSST photometric systems yield a smaller effect on the final counts than earlier assumptions on survey depths and footprints.

\begin{deluxetable*}{l c c c c c c}
    \tablecolumns{3}
    \label{tab:dp2depths}
    \tablecaption{DP2 DDF and SV 3k deg$^2$ depths}
    \tablehead{\colhead{Field} & \colhead{u} & \colhead{g} &\colhead{r} &\colhead{i} &\colhead{z} &\colhead{y} \\ \colhead{}  & \colhead{(mag)} & \colhead{(mag)} & \colhead{(mag)}& \colhead{(mag)}& \colhead{(mag)}& \colhead{(mag)}} 
    \startdata
        XMM\_LSS & 25.3 & \dots & \dots & \dots & \dots &\dots \\
        EDFS\_A & \dots & 25.3 & 25.2 & 25.0 & 24.1 & \dots \\
        EDFS\_B & \dots & 25.4 & 25.1 & 25.0 & 24.1 & \dots \\
        ECDFS & \dots & 25.8 & 25.5 & 25.3 & 24.7 &\dots \\
        ELAISS1 & 25.6 & 26.6 & 26.1 & 26.2 & 25.1 & 22.7 \\
        COSMOS & 26.0 & 26.8 & 26.3 & 26.1 & 25.3 & 23.9 \\
    \hline
        3k deg$^2$ (SV) & 24.2  &25.0  &24.9  &24.5 & 23.8 & 22.6 \\
    \enddata
    \tablecomments{Depths for the six DDFs and the SV 3k deg$^2$ are reproduced from Table 3 and Table 2 in \citet{RTN-011}, respectively.}
\end{deluxetable*}

For the DDFs and SV wide survey, we predict the number of potential brown dwarf detections using their pointings and estimated photometric depths (Table \ref{tab:dp2depths}). For the DDFs, we use the central coordinates and photometric depths reported in \citet{RTN-011}. We only consider the five DDFs with $z$ band photometry. For the SV Wide Area survey, we use the footprint of visit pointings related to the 3k deg$^2$, 750 deg$^2$, and 300 deg$^2$ surveys and the 3k deg$^2$ depths. We predict brown dwarf counts by determining the number of simulated objects that appear within each survey component's respective footprints and are brighter than the estimated depth limits. In doing so, we assume all pointings across the entire footprint are usable with good image quality and that the visits combine to reach the target coadded depths. These assumptions results in an overprediction from image quality and an underprediction from survey depth (i.e. the 750 deg$^2$ and 300 deg$^2$ SV components will reach $\sim$ 0.5 mag deeper in all three bands). Propagating uncertainties from the scatter in the absolute magnitude-spectral type relation used yields predicted count differences of $30\%$ for the DDF and SV fields. As such, our predicted counts should be viewed as an order of magnitude estimate, not an exact count.

Before discussing predicted detection counts in the DDFs and SV, we first confirm our simulation predicts the appropriate number of objects for DP1. The simulation predicts 19 brown dwarfs in ECDFS and 11 in EDFS brighter than the photometric depth limits in all three filters. These counts are exclusively for objects with spectral types L0 and later and do not reflect the M dwarfs selected in Section \ref{sec:candidates} nor do they predict the extendedness, saturation, or SNR of such objects. These predictions are in good agreement with our candidates and their phototemps from Section \ref{sec:candidates} (18 objects in ECDFS and 13 in EDFS with phototemps $\leq$ 2,300 K). The simulation was previously validated in \citet{Honaker25} against various age distributions and scale heights from kinematic samples of brown dwarfs \citep{dupuy17,Aganze22}.

\subsubsection{Deep Drilling Field Counts}
From the Ultracool Sheet, only one object fell within the DP1 footprints. There are six objects that fall within the Deep Drilling Field footprints: three in COSMOS (SDSS J100401.41+005354.9, L2; CFBDS J100113.05+022622.3, T5; CFBDS J095914.80+023655.2, T3.5), one in ELAISS1 (EROS-MP J0032-4405, L0), one in EDFS\_B (WISE J041358.14-475039.3, T9), and one in XMM\_LSS (2MASS J021857.92-061749.9, M8). Of the six objects, only 2MASS J021857.92-061749.9 and SDSS J100401.41+005354.9 have optical photometry included in the Ultracool Sheet suggesting that LSST may provide the first optical observations of the other four. Note that XMM\_LSS only includes $u$ band visits making characterization of new brown dwarf candidates difficult. 

To predict potential DDF brown dwarf detection counts, we compare the DDF footprints and estimated photometric depths (see \citealt{RTN-011}) to our synthetic population with $i,z,y$ band photometry. We consider an object detectable if it is within 1.8 $\deg$ from a DDF's central coordinate and brighter than at least one of the three magnitude limits. 

Our simulation shows the following detection counts. The number of objects that can be detected in both $i$ and $z$ and only in $z$ for each field is as follows: EDFS\_A: 34 and 21; EDFS\_B: 31 and 27; ECDFS: 44 and 34. As expected for brown dwarf SEDs, there is no case where an object is detected in the $i$ band but none of the redder bandpasses.
ELAISS1 and COSMOS have coverage in $i$, $z$, and $y$. The counts of objects that can be detected in $i$, $z$, and $y$, only in $i$ and $z$, and lastly only in $z$ are: ELAISS1: 37, 36, and 29; COSMOS 71, 28, and 17. For both the ELAISS1 and COSMOS DDFs, the $y$ band depth limit is respectively shallower than the $z$ band depth, resulting in no objects that can only be detected in $y$.

We predict there will be on the order of 400 objects detectable with at least one of the $i$, $z$, $y$ filters for the five DDFs considered ($\sim$ 90 objects/field) in DP2. Of those 400 objects, 281 would have sufficient photometry to construct at least one $i,z,y$-based color.
Our simulations also show that for all fields, the distribution of spectral types for the detectable objects greatly favor earlier spectral types (early- to mid-L dwarfs) but include some single and occasional two-band detections of later spectral types (mid-T dwarfs). Detection of later type dwarfs with multiple bands relies heavily on $y$ band photometry and occurs most often in the COSMOS field. The skew towards earlier spectral types, and therefore brighter objects that can be detected at greater distances, is a common shortcoming of magnitude-limited surveys and is not unexpected. The expected number of brown dwarf counts in the DDFs are significantly higher than the DP1 fields since the DDFs are $\sim$ 0.5 magnitudes deeper in each band but more importantly each DDF covers 1.8 deg$^2$ compared to the DP1 fields' $\sim$1 deg$^2$. We note that the predicted number of brown dwarfs detectable in the DDFs is much larger than the number of known objects (6) expected to appear within the fields: the DP2 DDFs will present an immediate opportunity to identify and characterize a substantial number of new  discoveries.

\subsubsection{Science Verification Survey Counts}
The DDF estimations show a significant, but manageable number of known brown dwarfs and potential new discoveries. While the DDFs are deeper than the SV survey footprint, the DDFs cover a total on-sky area of $\sim$ 9 deg$^2$, less than  1\% of the SV footprint. From the Ultracool Sheet, there are 396 known objects that fall within the 3k deg$^2$ survey footprint (110 M dwarfs, 188 L dwarfs, 88 T dwarfs, 5 Y dwarfs). While the late-T and Y dwarfs will most likely be undetectable by LSST unless they are exceptionally nearby, we anticipate the vast majority of the M and L dwarfs will be detectable. 

The $\sim$ 400 known objects within the SV footprint are a large, but manageable, number of objects to follow up and characterize. However, our simulation predicts that across the entire 3k deg$^2$ SV footprint, assuming the shallowest survey depths ($i,z,y=24.5,23.8,22.6$ mag), there will be $\sim 45,000$ objects detectable in at least one filter. Since only one filter is not enough information to separate a brown dwarf from other objects, we note that $\sim 33,000$ of the objects will be detected by both the $z$ and $y$ bands and $\sim 17,000$ of those objects will be detectable in all three bands. Notable brown dwarf archives like the Ultracool Sheet or SIMPLE archive \citep{simplearchive} each contain $\sim 4,000$ objects. Even the $17,000$ predicted objects with 3-band photometric coverage represent a 325\% increase. As such, brown dwarf science truly enters the realm of big data where careful thought must be put into follow-up observation strategies to maximize resource allocation.

\subsection{Future Outlook}
With our predictions for detection counts in DP2's Deep Drilling Fields and Science Verification surveys, we now turn to discuss two areas of importance for maximizing the value of LSST in DP2 and beyond.

\subsubsection{Infrared Cross-matching with LSST}
Since brown dwarfs emit the vast majority of their flux in the infrared beyond LSST's $y$ band, supplementing LSST with near- and mid-infrared measurements is necessary to fully characterize brown dwarf candidates. While this can be reasonably well done for LSST DP1 with existing surveys like VHS or CatWISE, LSST already has the capability to detect objects that are fainter than the respective infrared magnitude depths; as LSST progresses, only Euclid and Roman will be able to match LSST's depths. 

\begin{figure*}
    \centering
    \includegraphics[width=\textwidth]{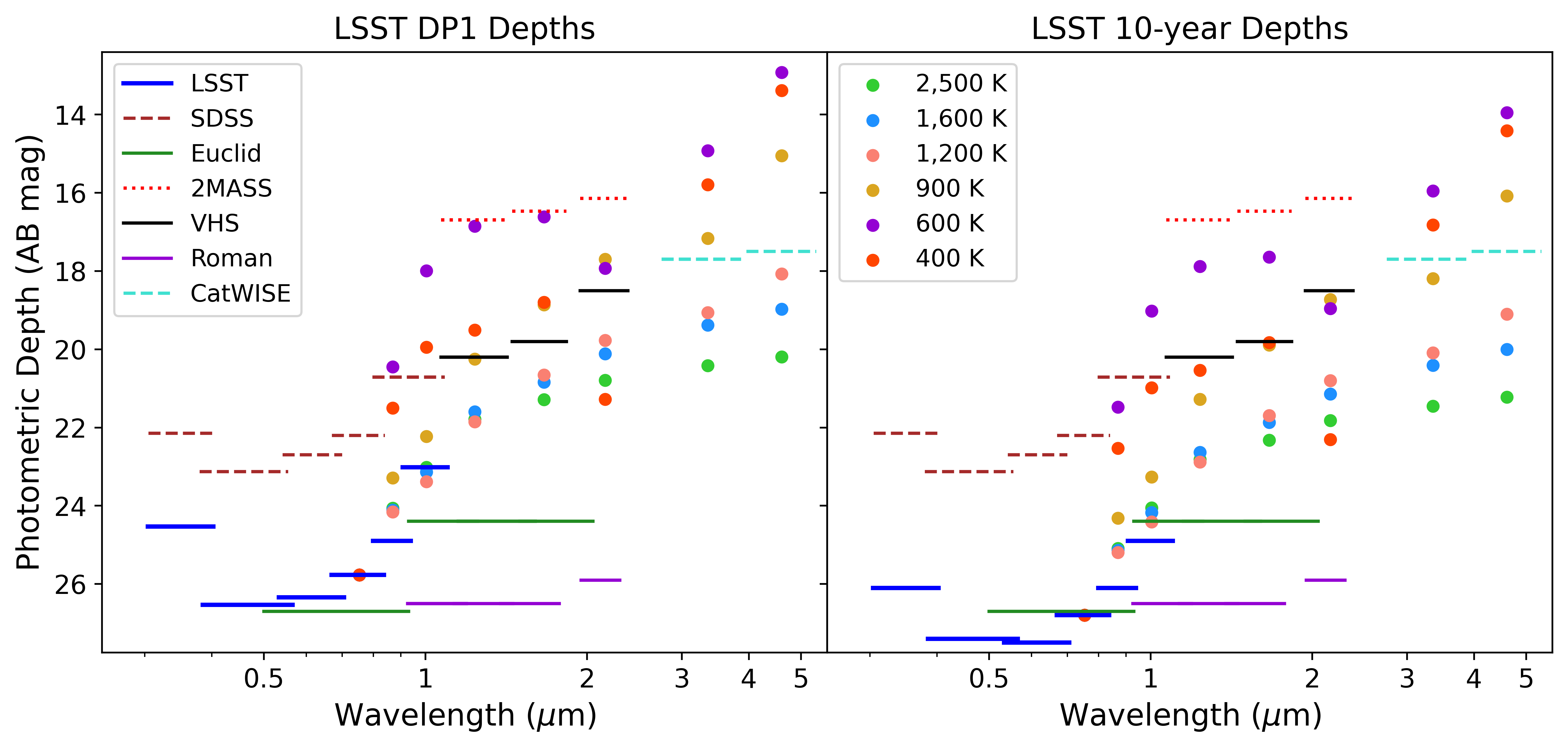}
    \caption{We simulated synthetic magnitudes in various infrared surveys for objects with temperatures between $T_{\text{eff}}= 400$ K to $2,500$ K that reach the LSST $i$ band photometric depth limit. Various survey wavelength coverage and depths in AB magnitudes are shown in both panels as horizontal lines. The simulated object magnitudes are shown as circles, color-coded by temperature. Surveys and objects retain the same color scheme for both panels. In the left panel, we flux calibrate objects to reach the DP1 $i$ band depth and nearly all objects are below 2MASS detection thresholds but above VHS and CatWISE limits. In the right panel, we repeat the process using the full LSST 10-year survey $i$ band depth and the majority of objects exceed the VHS faintness limits and are only detectable by Euclid or Roman.}
    \label{fig:depths}
\end{figure*}

In Figure \ref{fig:depths}, we compare the relative depths of different optical and infrared surveys to show which can adequately match LSST's depth. To do so, we model six objects with temperatures ranging from 2,500 K down to 400 K (spectral types $\sim$ M7 - Y2) using Sonora Diamondback (2,500 K, 1,600 K, 1,200 K, 900 K; \citealt{diamondback}) and  ExoRem (600 K, 400 K; \citealt{Baudino15,Charnay18,Blain21}) model atmospheres with solar metallicity and $\log{g} = 5$. We flux calibrate all objects to the $i$ band DP1 ($i=25.77$ AB mag) and 10-year ($i=26.8$ AB mag) survey median depths to simulate an object being barely detected in $i$ and dropping below detection thresholds in bluer bandpasses. Then, we calculate synthetic magnitudes in the MKO $J,H,K$ and WISE $W1,W2$ bandpasses. Already with LSST DP1 median depths ($r, i, z, y = 26.34, 25.77, 24.90, 23.02$ for ECDFS), brown dwarfs at the $i$ band detection limit are almost entirely below 2MASS detection limits and reach maximum VHS and CatWISE depths. 

In the right panel of Figure \ref{fig:depths}, we repeat the simulation using the estimated LSST 10-year depths. At full survey depths, objects that reach the $i$ band depth limit exceed VHS and CatWISE depths with the exception of only the coldest, most nearby objects. This effect is even greater for objects that drop out of the $i$ band and reach the $z$ band detection limit; in such cases, even DP1 depths exceed the equivalent VHS and CatWISE depths and full survey depths yield objects well below infrared survey depths. However, Euclid and Roman promise to complement LSST's depth with matching (and exceeding) infrared depth. These two space-based observatories will be indispensable to LSST UCD science.  As seen in Section \ref{sec:colorcuts}, joint optical and infrared colors are critical to ultracool dwarf identification and classification. In order to maximize the impact of LSST for brown dwarf science in future data releases, cross-matching must be done using surveys with appropriate depths. 

\subsubsection{LSST Astrometry}
As the LSST survey progresses, its astrometric precision promises to deliver high-quality parallaxes and proper motions. For astrometric purposes, LSST can be considered an extension of the Gaia mission for faint objects in the Southern sky. LSST uses the same GBDES astrometry algorithm that was used on the Dark Energy Survey \citep{gbdes}. However, reliable astrometric solutions are not expected to be available until after several data releases with a sufficient time baseline. Note that astrometric solution parameters will be available in data products before such time, but are not expected to be reliable, especially for faint, high proper motion objects. 

Eventually, LSST-only selection of UCDs may be possible, but will require proper motions. Historically, UCD searches have combined optical and infrared data, but UCD candidates can be distinguished without infrared information by their red optical colors, high proper motions, and large parallaxes. We originally attempted our candidate selection only using current LSST data, but found the results to be too inefficient and requiring inordinate amounts of follow up to confirm/exclude candidates' UCD nature. 

Reliable distances and kinematics for the unprecedented sample of UCDs that LSST will discover are critical for a multitude of science cases. Here we specifically discuss the luminosity function for UCDs. Using LSST parallaxes, one can construct a large volume-limited sample and measure the luminosity for UCDs, probing formation mechanisms, cooling rates, age distributions, binary fractions, and space densities. For the brown dwarf population synthesis model used in Section \ref{predictions}, we applied two different evolutionary models and produced two different synthetic populations. Reported predictions come from the population built using the Sonora Diamondback hybrid evolutionary models, but the same simulation can be built using the Sonora Bobcat cloudless evolutionary models. We compared the two populations as they would be detected by DP2 and found that significant differences in their luminosity functions can be identified, but only if objects have reliable distance measurements, which will not be the case for UCDs in DP2 or DR1. Despite low total sky coverage, a volume-limited sample of DP2 objects can probe different luminosity function morphologies related to the L/T transition (for a more detailed investigation of brown dwarf luminosity functions, see \citealt{Honaker25} and references therein). We attempted to build an optical color-based distance estimator using PS1 $z-y$ colors and absolute $z$ magnitudes from the Ultracool Sheet, but found the sharp turn between $0.85\lesssim (z-y) \lesssim1.1$ results in prohibitively large uncertainties. A distance estimator built from joint optical and infrared photometry may yield better results, but is beyond the scope of this work. 

LSST will produce exquisite astrometric measurements for objects, but reliable astrometric solutions for faint, high proper motion UCDs will require a substantial baseline extending several data releases into the survey. Yet, numerous science cases can immediately make use of DP2 and DR1 datasets if objects have reliable distance measurements. Consequently, cross-matching to adequately deep, existing infrared surveys will be a pivotal stopgap.

\section{Conclusion}\label{conclusion}

In this work, we have showcased LSST's potential for discovering and characterizing ultracool dwarfs (UCDs). Since LSST presents a new optical photometric system, we began by searching for known UCDs within Data Preview 1 (DP1) and found one known object from the Ultracool Sheet. Using UCD candidates from the Dark Energy Survey, we recovered 17 known UCDs across 3 DP1 fields and spectral typed 9 of them using Euclid spectroscopy. We repeated the search process for known low-mass stars and UCDs from the Gaia Catalog of Nearby stars and found 40 objects with LSST counterparts. However, the majority of nearby, low-mass stars with absolute Gaia $G$ magnitude $M_G> 12$ saturate in LSST's 30 second exposures and create artifacts that can easily be mistaken for color outliers. LSST will be a powerful tool for understanding the low-mass end of the main sequence, but can be limited for very nearby ($< 100$ pc) F, G, or K stars. 

Next, we leveraged the synergies between LSST and Euclid to search for new UCD candidates in the ECDFS and EDFS fields. Through a series of quality, signal-to-noise ratio, extendedness, and color cuts, we found 89 UCD candidates across the two fields, including 17 new discoveries and recovering 72 objects recently identified in \citet{Zerjal}, \citet{Zhang25}, and \citet{dalponte23}. In constructing our candidate sample, we detailed a more robust approach to select point sources where the LSST pipelines struggle with faint object star/galaxy separation. We also investigated the efficiency of LSST-only, Euclid-only, and joint colors at separating UCD candidates from other point sources. We found that colors using the LSST $i$ or Euclid $I_E$ bands provided the most separation. To characterize our candidates, we developed a photometric temperature estimator (phototemp) using the SANDee atmospheric model grid; 58 of the candidates have phototemps between 2,400 K and 2,800 K, consistent with M dwarfs. The remaining 31 candidates have phototemps ranging from 2,100 K to 2,300 K, roughly consistent with L dwarfs. Phototype methods are well suited to large surveys like LSST, but require substantial empirical samples of known objects to effectively classify unknown sources.

Finally, we used synthetic populations of brown dwarfs to predict estimated counts for upcoming LSST Data Preview 2 (DP2). We expect DP2 to contain several hundred known UCDs across the Deep Drilling Fields (DDFs) and Science Verification (SV) survey footprints. These known objects will be critical to better understanding the LSST photometric system as well as building a database of objects with known colors. In additional to known objects, our simulation predicts DP2 has the potential to detect $\sim 45,000$ brown dwarfs in at least one band, and $\sim 17,000$ may be detected in $i,z,y$ in the SV areas. Several hundred new UCDs may be discovered in the DDFs. These discoveries would increase the current sample of brown dwarfs by a full order of magnitude but require coordinated community follow-up since that many objects will significantly outpace available observational resources. We also highlighted some key challenges in leveraging such a dataset; proper care must be taken to cross-match using appropriately deep infrared surveys to fully leverage LSST's capabilities. Astrometrically, UCDs must rely on external distance and proper motion measurements until later in the LSST survey, but such measurements are critical to many science cases.

LSST will be a generation-defining survey that requires the entire Rubin community's collaborative efforts to maximize scientific gain. For UCDs, LSST's wide and deep coverage, alongside the built-in synergies with Euclid and Roman, will be instrumental in addressing some of the field's open questions. We have showcased a small piece of LSST's capabilities and included the lessons we have learned as we begin to understand the survey's data products, photometric system, and analysis. In sharing these lessons with the broader scientific community, we can better prepare for the survey's imminent beginning and upcoming data releases.

\facilities{Rubin:Simonyi (LSSTComCam), Gaia, Euclid}

\software{\texttt{Astropy} \citep{astropy2013,astropy2018,astropy2022}, \texttt{TOPCAT} \citep{Topcat}, \texttt{Matplotlib} \citep{matplotlib}, \texttt{NumPy} \citep{numpy}, \texttt{SPLAT} \citep{splat}, \texttt{LSDB} \citep{lsdbBASE}}

\begin{acknowledgments}
    We thank the Rubin Observatory staff and community for their hard work bringing the survey to life. We also thank the observatory for the computational resources and tutorials available through the Rubin Science Platform. This work benefited from discussions with the Stars, Milky Way, and Local Volume collaboration and the Solar Neighborhood working group and we thank both organizations. The authors thank the anonymous referee for comments and suggestions. EJH is funded by Delaware Space Grant College and Fellowship Program (NASA Grant 80NSSC25M7071) and thanks fellow graduate students at UD for insightful discussions and feedback on visual elements in this work. MZ and ELM are supported by the European Union (ERC, SUBSTELLAR, project number 101054354). FBB acknowledges support from NSF Awards 2511639 and 2308016. SC is grateful for the support received from the University of Delaware Doctoral Fellowship of Excellence, and the NASA FINESST program, Grant 80NSSC25K0312.
    
    This research has made use of the SVO Filter Profile Service ``Carlos Rodrigo'', funded by MCIN/AEI/10.13039/501100011033/ through grant PID2023-146210NB-I00. This work has made use of the SIMPLE Archive of low-mass stars, brown dwarfs, and directly imaged exoplanets: 10.5281/zenodo.13937301. This work has made use of the Euclid Quick Release Q1 data from the Euclid mission of the European Space Agency (ESA), 2025, https://doi.org/10.57780/esa-2853f3b. This work has benefited from The UltracoolSheet at http://bit.ly/UltracoolSheet, maintained by Will Best, Trent Dupuy, Michael Liu, Aniket Sanghi, Rob Siverd, and Zhoujian Zhang, and developed from compilations by \citet{DupuyLiu12}, \citet{DupuyKraus13}, \citet{Deacon14}, \citet{Liu16}, \citet{Best18}, \citet{Best21}, \citet{Sanghi23}, and \citet{Schneider23}.

\end{acknowledgments}
\appendix

\section{Phototemp determination}\label{phototemp_details}
In Section \ref{sec:phototemp}, we classify candidate UCDs using a phototype method similar to \citet{skrzypek15} using the SANDee models as templates and the candidates' LSST and Euclid photometry. We anchor this process using the candidates' measured magnitudes and the template colors. To summarize, each candidate has photometry from LSST $i,z,y$ and Euclid $Y_E,J_E,H_E$, which together comprise $N=6$ bands and create the vector of magnitudes $\hat{m}_b$, where $b$ counts each of the six bands. Note that we avoid using the broad Euclid $I_E$ band and instead opt for the narrower LSST $i,z,y$ bands to capture smaller changes between templates in the optical. Colors for the candidates and templates are calculated with respect to a single band denoted as the reference band, $B$. Thus, any given color is $b-B$ for each band, $b$, and the entire system is anchored by the reference band such that its color with itself is 0 (i.e. if one uses $J_E$ as the reference band, all colors are computed with respect to $J_E$ and $J_E-J_E=0$). Note this approach only uses colors containing the reference band. All of our SANDee templates have colors $c_{b,t}$ where the $t$ index counts the template.  For each candidate, we compute a reference magnitude, 
\begin{equation}
  \hat{m}_{B,t} = \frac{\sum_{b=1}^{N} (\hat{m}_b - c_{b,t})/\sigma_b^2}{\sum_{b=1}^{N}(1/\sigma_b^2)} ,
\end{equation}
where $\sigma_b$ is the candidate's measured uncertainty in each band, $b$. Including photometric uncertainty weights the estimated reference magnitudes by the quality of measurement so that magnitudes with large uncertainties do not dominate the phototype estimation. In practice, for a system anchored using the $J_E$ band, calculating $\hat{m}_{B,t}$ is the same as estimating a candidate's $J_E$ magnitude using $N_b$ bands given a specific template's colors $c_{b,t}$. Next, to assign a template to the candidate, we compute and minimize the $\chi^2$ value for each template:
\begin{equation}
    \chi^2 = \sum_{b=1}^{N} \left(\frac{\hat{m}_b - \hat{m}_{B,t} - c_{b,t}}{\sigma_b}\right)^2.
\end{equation}
The template with the lowest $\chi^2$ is then assigned as the best match and phototype. However, since we are not using a traditional (empirical) set of templates labeled by spectral type but rather a set of atmospheric grids sorted by temperature, we dub the best-fitting template as the candidate's ``phototemp''.

For consistency with previous studies, we anchor our phototemp system using the $J_E$ band, but the choice of reference band is very important and has drastic impact on the classification. Here, we evaluate this impact and make suggestions for future studies. We compute $\chi^2$ values using colors anchored with each of the seven bands (i.e. we used each band as the reference band $m_B$) and compared the $\chi^2$ values for all candidates. Anchoring with the $i$ band resulted in the highest $\chi^2$ values across the board while using the $H_E$ band yielded the lowest values. The second lowest $\chi^2$ come from using the $J_E$ band as the reference band. Resulting phototemps from using $J_E$ versus $H_E$ agree for 80/89 candidates while for the remaining  candidates the $J_E$-based phototemp is within 100 K (one template) of the $H_E$-based equivalent (in 8 of 9 cases, the $J_E$ phototemp is 100 K cooler). Historically, phototype systems have been anchored using whichever band is best constrained, generally a $J$ band equivalent. In our case, both $J_E$ and $H_E$ are equally well-constrained since we imposed the requirement that $\mathrm{SNR}_{J_E,H_E}\geq3$. Our choice of using $J_E$ should be revisited in the future once large samples of LSST-confirmed UCDs are available. While anchoring using optical $z$ bands is generally disfavored due to high dependencies on the bandpass red cutoff shape, incorporating additional optical bands (e.g. $g,r$) with LSSTCam may enhance optical phototype estimators.  

\bibliography{sample701}{}
\bibliographystyle{aasjournalv7}

\end{document}